\documentclass[%
 reprint,
superscriptaddress,
nofootinbib,
nobibnotes,
 amsmath,amssymb,
 aps,
]{revtex4-2}

\usepackage{graphicx}
\usepackage[utf8]{inputenc}
\usepackage[T1]{fontenc}
\usepackage{float}  
\usepackage{wrapfig}
\usepackage{xr}
\usepackage{hyperref}
\usepackage{dsfont}
\usepackage{dcolumn}
\usepackage{bm}
\usepackage{comment}
\usepackage{natmove}
\usepackage{dsfont}
\usepackage{xcolor}
\usepackage{enumitem}
\usepackage[nolist,nohyperlinks]{acronym}

\begin{document}
\title{Reaction-driven Diffusiophoresis of Liquid Condensates: \\Mechanisms for Intra-cellular Organization}
\author{Gregor Häfner}
    \affiliation{Georg-August Universität Göttingen, Institut für Theoretische Physik, Friedrich-Hund~Platz~1, 37077 Göttingen, Germany}%
    \affiliation{Max Planck School Matter to Life, Jahnstraße 29, 69120 Heidelberg, Germany}
\author{Marcus Müller}
    \email{mmueller@theorie.physik.uni-goettingen.de}
    \affiliation{Georg-August Universität Göttingen, Institut für Theoretische Physik, Friedrich-Hund~Platz~1, 37077 Göttingen, Germany}%


\date{\today}%
\begin{abstract}
The cellular environment, characterized by its intricate composition and spatial organization, hosts a variety of organelles, ranging from membrane-bound ones to membraneless structures that are formed through liquid-liquid phase separation. Cells show precise control over the position of such condensates. We demonstrate that organelle movement in external concentration gradients, \textit{diffusiophoresis}, is distinct from the one of colloids because fluxes can remain finite inside the liquid-phase droplets and movement of the latter arises from incompressibility.
Within cellular domains diffusiophoresis naturally arises from biochemical reactions that are driven by a chemical fuel and produce waste. Simulations and analytical arguments within a minimal model of reaction-driven phase separation reveal that the directed movement stems from two contributions:
Fuel and waste are refilled or extracted locally, resulting in concentration gradients, which (i) induce product fluxes \textit{via} incompressibility and (ii) result in an asymmetric forward reaction in the droplet’s surroundings (as well as asymmetric backward reaction inside the droplet), thereby shifting the droplet's position. We show that the former contribution dominates and sets the direction of the movement, away from or towards fuel source and waste sink, depending on the product molecules' affinity towards fuel and waste, respectively. The mechanism thus provides a simple means to organize condensates with different composition. Particle-based simulations and systems with more complex reaction cycles corroborate the robustness and universality of this mechanism.
\end{abstract}

\maketitle

\begin{acronym}[MPTPS]
  \acro{1D}{one-dimensional}
  \acro{2D}{two-dimensional}
  \acro{3D}{three-dimensional}
  \acro{CUDA}{Compute Unified Device Architecture by Nvidia\textregistered}
  \acro{HDF5}{Hierarchical Data Format version 5}
  \acro{GPU}{graphics processing unit}
  \acro{HKCA}{Hoshen-Kopelman cluster analysis}
  \acro{RDA}{reaction-driven assembly}
  \acro{SCMF}{single-chain-in-mean-field}
  \acro{SI}{supporting information}
  \acro{JSC}{J\"ulich Supercomputing Centre}
  \acro{DFG}{Deutsche Forschungs Gemeinschaft}
  \acro{SOMA}{SOft coarse-grained Monte carlo Acceleration}
  \acro{UDM}{Uneyama-Doi model}
  \acro{XML}{Extensible Markup Language}
\end{acronym}

\newcommand{\Nbar}{\ensuremath{\bar{\mathcal N}}}
\newcommand{\R}{\ensuremath{R_{\textrm e}}}
\newcommand{\kT}{k_{\textrm B}T}
\newcommand{\eg}{\textit{e.\,g.}}
\newcommand{\ie}{\textit{i.\,e.}}

\newcommand{\GH}[1]{{\color{teal}GH: #1}}
\newcommand{\MM}[1]{{\color{blue}MM: #1}}


The cellular environment, in which biochemical processes occur, is characterized by diverse local compositions with a high degree of complexity and spatial organization. There exists a multitude of organelles, which can be membrane-bound, such as mitochondria and lysosomes, \cite{heald_morphology_2014} and membrane-less, such as centrosomes, P granules or cajal bodies. \cite{uversky_intrinsically_2017,gomes_molecular_2019,choi_physical_2020} The latter are typically formed by liquid-liquid phase separation \cite{brangwynne_germline_2009,brangwynne_active_2011,hyman_liquid-liquid_2014,berry_physical_2018}. The spatial organization is integral to cells by fostering biochemical reaction cycles and orchestrating critical functions within cells, such as RNA metabolism and signaling \cite{banani_biomolecular_2017,nakashima_biomolecular_2019,lyon_framework_2021,spruijt_open_2023}. Understanding how these organelles are positioned and transported within the cellular milieu is a central question in cell biology. \cite{saha_polar_2016,mogre_getting_2020}.
Typically, directed transport processes are thought of to rely on motor proteins or cytoplasmic flows. \cite{mogre_getting_2020}. Recently, simpler transport mechanisms that do not rely on these effects have gained attention. For instance, diffusiophoresis, which describes the movement of hard colloids in an external concentration gradient \cite{anderson_diffusiophoresis_1984,anderson_transport_1986}, has been hypothesized to be also implicated in transport of intracellular condensates \cite{sear_diffusiophoresis_2019} and is indeed observable for peptide and DNA condensates in salt concentration gradients \cite{doan_diffusiophoresis_2023}. However, biological condensates rarely are hard, impenetrable objects, but rather interact softly with their surrounding, including external diffusion and fluxes through condensates. It is hence necessary to extend our understanding of such concepts to liquid-liquid phase separation.  

Reducing the complexity of biological systems experimentally and theoretically has proven a valuable tool to understand the underlying physics. Complex coacervates formed by charged polymers or proteins can mimic the fluid-like properties of membraneless organelles and can be engineered to involve or catalyze chemical reactions \cite{drobot_compartmentalised_2018,nakashima_reversible_2018,donau_active_2020,spruijt_open_2023,dai_engineering_2023}. Typically, their involvement in biochemical reactions keeps the organelles out of equilibrium, aiding to control their size and to suppress coalescence. \cite{zwicker_centrosomes_2014,nakashima_active_2021} Theoretically, these phenomena can be explained by the use of minimal models that feature phase separation in combination with simple reaction cycles. For instance, in a phase separating system, whose components can switch between hydrophilic and hydrophobic by driven reactions, coalescence arrests at a finite size of droplets, introducing a length scale. \cite{glotzer_reaction-controlled_1995,christensen_phase_1996,zwicker_suppression_2015} Chemical reaction cycles further have been demonstrated to influence nucleation dynamics \cite{cho_tuning_2023,ziethen_nucleation_2023} or can even divide condensates. \cite{zwicker_growth_2017,golestanian_division_2017,weber_physics_2019-1,bauermann_energy_2022} Catalytic properties for reactions of organelles in finite domains have also been shown to introduce positioning and self-propulsion. \cite{demarchi_enzyme-enriched_2023}

In this work, we demonstrate a fundamental mechanism to direct organelles to or away from the center of a finite domain, such as the cell lumen. We perform simulations with two complementary levels of description, a continuum model as well as a particle-based one. 
To understand the basic principles of diffusiophoresis of liquid phases, we start with a passive droplet (condensate) in externally maintained concentration gradients of two solvents. One has a strong repulsion to the droplet molecules and one has varying interactions. Depending on the latter, there is a flux of the second solvent through the droplet that leads to its motion by virtue of the incompressibility of the two solvents and the droplet liquid. Furthermore, we describe a minimal model, which involves a reaction cycle for molecules that dissolve in their precursor state, but phase separate from solution in their product state, termed \ac{RDA}. To push the system out of equilibrium, the forward reaction is chemically driven, using a fuel to enable the reaction and creating waste in the process. Without exchange of chemical components with the surroundings, the system will convert all fuel and equilibrate into a homogeneous state. To avoid this and create a statistically stationary state out of equilibrium, fuel sources and waste sinks need to be introduced. In a biological context, these can arise from channels and activity at the cell membrane, whereas in synthetic systems, they can be introduced by a fluorinated oil phase that surrounds a microfluidic droplet, as recently demonstrated by Bergmann \textit{et al.}. \cite{bergmann_liquid_2023} The sinks and sources give rise to fluxes of fuel and waste. By accurately treating the incompressibility of the system, we demonstrate that this induces a product flux in opposite direction (directing the motion of condensates). The direction of transport crucially depends on the interaction of product with fuel and waste. Thus this provides a simple means to grow and direct chemically fueled aggregates in a controlled manner. Analytically, we demonstrate that two additional, subdominant contributions to the movement arise from an asymmetric production in the condensates' surroundings as well as an asymmetric reversion of product on the inside.
Finally, we illustrate the robustness of the results for a selection of scenarios: Combining passive and \ac{RDA}-formed droplets, we show that the mechanism allows to simultaneously move droplets in opposite direction. Additionally, we consider a system, where product is amphiphilic and self-assembles into micelles or vesicles. Here the favorable interactions of the hydrophilic head groups with fuel suffice to guide their growth and direct the aggregates' motion.

\section{Results and Discussion}
\label{sec:results}

We consider a solution, consisting to a majority of solvent $S$, in which components undergo a simple reaction cycle. In a forward reaction, a hydrophilic reactant $R$, acting as a precursor, reacts with a fuel $F$, a high-energy molecule. The reaction creates a hydrophobic product $P$, that phase separates from solution, and some waste $W$, \ie,  $R + F \rightarrow P + W$ with forward reaction rate $r_f$. The reaction cycle is completed by a spontaneous reversion of the product molecule to precursor, $P\rightarrow R$, at reaction rate $r_b$. 

We perform continuum model simulations where the system state is fully characterized by the local concentration fields, $\phi_c$ for $c=P,R,S,F,W$. The system's equilibrium is governed by the Flory-Huggins-de Gennes free-energy functional, \cite{de_gennes_dynamics_1980} 
\begin{align}
	\frac{\mathcal{F}[\{\phi_c\}]}{\sqrt{\Nbar}\kT} =  \int \frac{d\mathbf{r}}{\R^3} \sum_c &\left(\frac{N_P}{N_c} \phi_c \ln \phi_c + \frac{\R^2}{24\phi_c} \left|\nabla \phi_c\right|^2 \right. \nonumber \\
 &\left. + \frac{1}{2}\sum_{c'\neq c} \chi_{cc'}N_P \phi_c\phi_{c'} \right) \nonumber \\
 &+ \pi(\mathbf{r})N_P \left(\sum_c\phi_c-1\right)
	\label{eq:free-energy-functional}
\end{align}
The mean end-to-end distance, $\R$, of the product molecule sets the length scale, and we measure all energies in units of $\kT$, with $k_{\text{B}}$ being the Boltzmann constant and $T$ the temperature, respectively.
$N_P=N_R$ denote the chain discretization or molecular volume of the product or reactant polymer, and $\Nbar$ is the invariant degree of polymerization, which dictates the strength of thermal fluctuations. $N_c$ with $c=S,\, F$, and $W$ denote the molecular volumes of solvent, fuel, and waste. The binary interactions are determined by the Flory-Huggins parameter, $\chi_{cc'}$ and the system is considered incompressible, which is enforced by the Lagrange multiplier, $\pi(\mathbf{r})$.

The dynamics of the concentration fields account for diffusive equilibration and driven, chemical reactions,
\begin{align}
	\frac{\partial\phi_c(\mathbf{r},t)}{\partial t} = \nabla \cdot \left[ \Lambda_c(\mathbf{r},t) \nabla \mu_c(\mathbf{r},t) \right] + s_{c}(\mathbf{r},t).
    \label{eq:time-evolution}	
\end{align}
The first term minimizes the free energy, locally conserving the concentration. Here gradients of the chemical potentials, $\mu_c(\mathbf{r})=\frac{\delta\mathcal{F}}{\delta \phi(\mathbf{r})}$, give rise to mass fluxes, $\mathbf{j}_c = - \Lambda_c \nabla \mu_c$. The Onsager coefficient $\Lambda_c$ is proportional to the diffusion coefficient, $D=\lambda \R^2$ which defines a time scale $\lambda^{-1}=\frac{\R^2}{D}$. Thermal noise is included as noise in the flux, see \autoref{eq:full-time-evolution}. The second term in the time-evolution equation, $s_c(\mathbf{r})$, describes the chemical reactions, which are considered to follow simple rate equations.
The model and parameters are described in more detail in the \autoref{sec:theretical_framework} section.

Although the parameters of the Flory-Huggins-de Gennes free-energy functional are inspired by polymer physics, the free-energy functional and the kinetic equations describe the universal features of macrophase separation coupled to reactions and, thus, are applicable to a wide variety of systems. \cite{weber_physics_2019-1,zwicker_evolved_2022} In this work we do not consider effects that arise from hydrodynamics, which could be incorporated using multi-fluid continuum models, \cite{doi_dynamic_1992} or multi-particle collision dynamics. \cite{malevanets_mesoscopic_1999,gompper_multi-particle_2009} This is valid for the slow transport of droplets in a highly viscous environment.

To illustrate the relevant effects, we consider four scenarios with increasing complexity: (i) We start with a phase-separating but chemically inactive system subjected to an externally maintained flux of solvents. (ii) We continue with chemically active systems, where droplets form by \ac{RDA}, and the continuous turnover of fuel to waste and the concomitant local refilling of the former naturally gives rise to concentration gradients. Here, either waste or fuel are only implicitly represented and treated to be part of the solvent. (iii) Then, both components are treated explicitly, and (iv) we show an exemplary scenario where the product is amphiphilic and assembles into micelles or vesicles that are also transported.

\subsection{Passive Droplets in External Concentration Gradients}
\label{sec:passive_diffusiophoresis}

\begin{figure}
    \centering
    \includegraphics[width=0.48\textwidth]{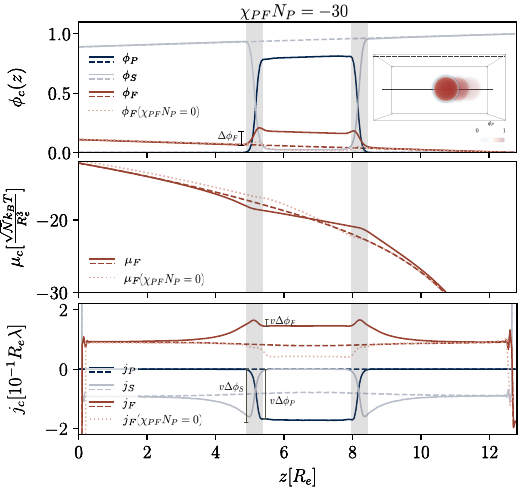}
    \caption{Exemplary profiles, $0\leq z \leq L_z/2$, of a passive droplet in external concentration gradients for preferential interactions with the $F$-solvent, $\chi_{PF}N_P=-30$, at time $t\lambda=20$. Concentration profiles (top), chemical potential of $F$-solvent (middle) and fluxes (bottom) plotted along two lines, one that passes by the droplet (dashed) and one that runs through its center (solid). In the bottom panel, the flux jumps, $v \Delta \phi_c$ with $v=-0.21\R\lambda$, are indicated for each component. Example snapshots of the \ac{3D}-morphology are overlayed in the inset of the top panel, and the positions of the line profiles -- inside and outside of the droplet -- are indicated. 
    Additionally, each panel shows the respective quantity for the $F$-solvent for the neutral interaction, $\chi_{PF}N_P=0$ at time $t\lambda=62$, as dashed pink lines. Notice that the reference (far-field, outside the droplet) chemical potential is slightly shifted compared to that of the system with favorable interactions, but not shown here.}
    \label{fig:inactive_test_profiles}
\end{figure}

To introduce the basic concept of diffusiophoresis for phase-separated condensates, let us begin with passive droplets in externally maintained concentration gradients but without chemical reactions. For this system, we only require three components, the droplet material, $P$, and two solvents, $S$ and $F$. The first one is highly repulsive towards the droplet material $\chi_{PS}N_P=30$, giving rise to liquid-liquid phase separation, whereas we vary the interactions of the second solvent, $F$, with the droplet material, $\chi_{PF}N_P=-30,\,0$, and $30$. In the remainder of this work, $F$ will take the role of a chemical fuel that enables the forward reaction, but it is purely passive in this section. 

A concentration gradient is maintained by introducing local sources and sinks. At the sides of the domain $z=0$, we place an $F$-source and $S$-sink, $S\rightarrow F$, while in the center slab, $z=L_z/2=12.8\R$, the opposite happens, $F\rightarrow S$, with higher rate constant. Starting without droplet material in the system, stationary concentration profiles build up with $\phi_F(z=0)\approx 0.1$ and approximately linearly decreasing towards the center, $\phi(z={L_z}/{2})\approx 0$. In this steady state, there is a continuous flux of $F$-solvent from the side, $z=0$, into the center, $z=L_z/2$, and a flux of $S$-solvent in opposite direction. Next, we replace the $S$-solvent by droplet material, $P$, inside a sphere of radius $R\approx 1.4 R_e$ slightly off centered in the domain. The droplet instantly equilibrates locally and interfaces build up. Since $S$-solvent is strongly repelled from the droplet phase, its flux inside almost vanishes. The $F$-solvent on the other hand, may have nonpreferential or attractive interactions towards the droplet material. In these scenarios, it has a finite concentration inside the droplet and also diffuses through it. Incompressibility demands the total concentration be constant and hence the total flux vanishes, $\sum_c \mathbf{j}_c = 0$.\footnote{Continuity equation and incompressibility assert that the divergence of the total flux vanishes. Due to cylindrical symmetry around the axis along the gradient of the external concentration through the center of the droplet, there is no rotational component to the flux, \ie, the flux must be constant, and this constant can be set to $0$ by choice of the reference frame.} This, in turn, couples the finite fuel flux, $\mathbf{j}_F$, inside the droplet to the flux of droplet material, $\mathbf{j}_P$. Outside the droplet, $\phi_P\approx 0$, we have $\mathbf{j}_F \approx -\mathbf{j}_S$, whereas inside the droplet, $\phi_S\approx0$, the flux of $F$-solvent cancels the flux of droplet material, $\mathbf{j}_P\approx-\mathbf{j}_F$. Thus, the passive droplet, $P$, moves in the opposite direction than the $P$-attractive $F$-solvent, \ie, towards the side of the simulation cell. 

\autoref{fig:inactive_test_profiles} shows example profiles for the attractive scenario, $\chi_{PF}N_P=-30$, giving concentrations, chemical potentials and resulting fluxes for a cut through the system that crosses (solid) and excludes (dashed) the droplet, respectively. Clearly, for the attractive scenario, $F$-solvent is enriched inside the droplet. The interfaces are \textit{locally} in equilibrium, \ie, the chemical potentials are continuous across the narrow interface, as shown exemplary for $\mu_F$. Under the assumption of a sharp interface, its movement in $z$-direction demands a discontinuity of the $F$-flux, $\Delta j_{F,z} = v\Delta\phi_F$, which is indicated in the figure, where $v$ denotes the velocity of the droplet. Likewise, there must be an analog interfacial discontinuity, $\Delta j_{P,z} = v\Delta\phi_P$, in the $P$-flux and in the $S$-flux. In order to fulfill the requirement, $\Delta j_{F,z}/\Delta\phi_F=\Delta j_{P,z}/\Delta\phi_P=\Delta j_{S,z}/\Delta\phi_S$ at the interface, the concentration profiles around the droplet's interface are altered, resulting in a smaller chemical potential gradient of $F$ is inside of the droplet and larger one in front of the interface, outside of the droplet, see middle panel of \autoref{fig:inactive_test_profiles}. The concomitant fluxes along $z$-direction are depicted in the bottom panel. By virtue of incompressibility, $j_{P,z}\approx-j_{F,z}$ and $j_{S,z}\approx 0$ inside of the droplet. The panel also indicates the three interfacial jump conditions, $\Delta j_{c,z}=v \Delta\phi_c$ for $c=P,\, F$, and $S$, at the droplet's interface which are blurred by the finite interface width.

\begin{figure}
    \centering
    \includegraphics[width=0.48\textwidth]{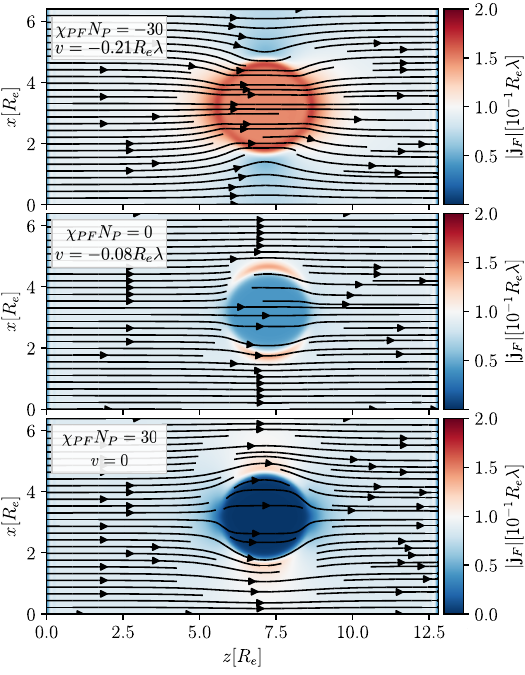}
    \caption{$F$-solvent fluxes plotting the direction as arrows and the absolute values colored underneath for the different interaction scenarios: (top) $\chi_{PF}N_P=-30$ at $t\lambda=20$, (middle) $\chi_{PF}N_P=0$ at $t\lambda=62$, (bottom) $\chi_{PF}N_P=30$ in the stationary state, where the droplet remains at position it was set up on. Shown is a slice through the center of the $y$-dimension.}
    \label{fig:inactive_test_fluxes2d}
\end{figure}

The one-dimensional flux profiles in \autoref{fig:inactive_test_profiles} only provide an incomplete description (also see \autoref{sec:1DSI} of the \ac{SI}), and the cylinder-symmetrical $F$-solvent flux around the droplet is illustrated in \autoref{fig:inactive_test_fluxes2d}.

For an attraction, $\chi_{PF}N_P=-30$, between $F$-solvent and droplet material $P$, $F$ preferentially diffuses through the droplet and the mass flux is focused into the droplet. This, in turn, increases the $P$-flux in opposite direction, $j_{P,z}\approx-j_{F,z}$, and quickly moves the droplet towards the $F$-source, at a velocity $v = -0.21\R\lambda$.

In the opposite case, if there is a strong repulsion towards the droplet material, $\chi_{PF}N_P=\chi_{PS}N_P=30$, the concentration of $F$-solvent inside of the droplet is vanishingly small, just as the $S$-solvent concentration is. Both solvents preferentially avoid the droplet and flow past it. In this case, the droplet does not move at all, $v=0$.

In the intermediate case of nonpreferential interactions, $\chi_{PF}N_P=0$, our choice of a higher molecular volume of $P$, $N_P=10N_F$ leads to an entropy-induced reduction of $F$-solvent concentration inside of the droplet compared to the immediate outside (as indicated by the dotted line in the upper panel of \autoref{fig:inactive_test_profiles}).  This leads to a de-focusing of the $F$-flux around the droplet. The flux inside the droplet is reduced and, therefore, also the droplet velocity is smaller compared to the $P$-$F$-attractive scenario, $v=-0.08\R\lambda$. The $F$-solvent concentration, chemical potential, and flux are plotted as dotted lines in \autoref{fig:inactive_test_profiles} for comparison. 

The measured droplet velocity can be related to fluxes and concentrations, as worked out in \autoref{sec:velocities_analytical} of the \ac{SI} for the general \ac{RDA} case. In the limit of no reactions and small variation of concentration profiles across the droplet volume, the comparison reduces to $v=-\frac{j_F(\mathbf{r}_\mathrm{cm})}{\phi_P(\mathbf{r}_\mathrm{cm})}$. This comparison is demonstrated in \autoref{sec:velocity_passive_droplet} of the \ac{SI}.

Summarizing this section, external fluxes can move passive, phase-separated droplets by diffusive solvent fluxes through these droplets. The flux of droplet material couples to the solvent flux \textit{via} incompressibility. In this way the droplet is transported towards the source of the more favorable solvent. The droplet velocity is the higher, the more enriched the solvent is inside of the droplet. This concept will reappear in the chemically active systems with \ac{RDA}-formed droplets in the next sections.

\subsection{RDA-formed Droplets with Implicit Waste or Fuel}

In the following, we consider reactions, where the droplet material, product $P$, is continuously turned over into a precursor state, reactant $R$, that is hydrophilic and expelled from the droplet. The precursor, $R$, can be transformed into the product, $P$ by a reaction with fuel, $F$, which is converted to waste, $W$, in the reaction. Hence, the whole system acts as a fuel sink and a waste source. In a first step, we treat one of the latter components implicitly, as part of the solvent.

\subsubsection{Implicit Waste}
\label{sec:implicit_waste}
When treating waste implicitly, we only consider the four components $P,R,S$, and $F$, and lump the waste together with the solvent. Hence, the forward reaction becomes $R+F\rightarrow P+S$, and the fuel source simultaneously acts as the waste (solvent) sink. 

\begin{figure}
	\centering
	\includegraphics[width=0.48\textwidth]{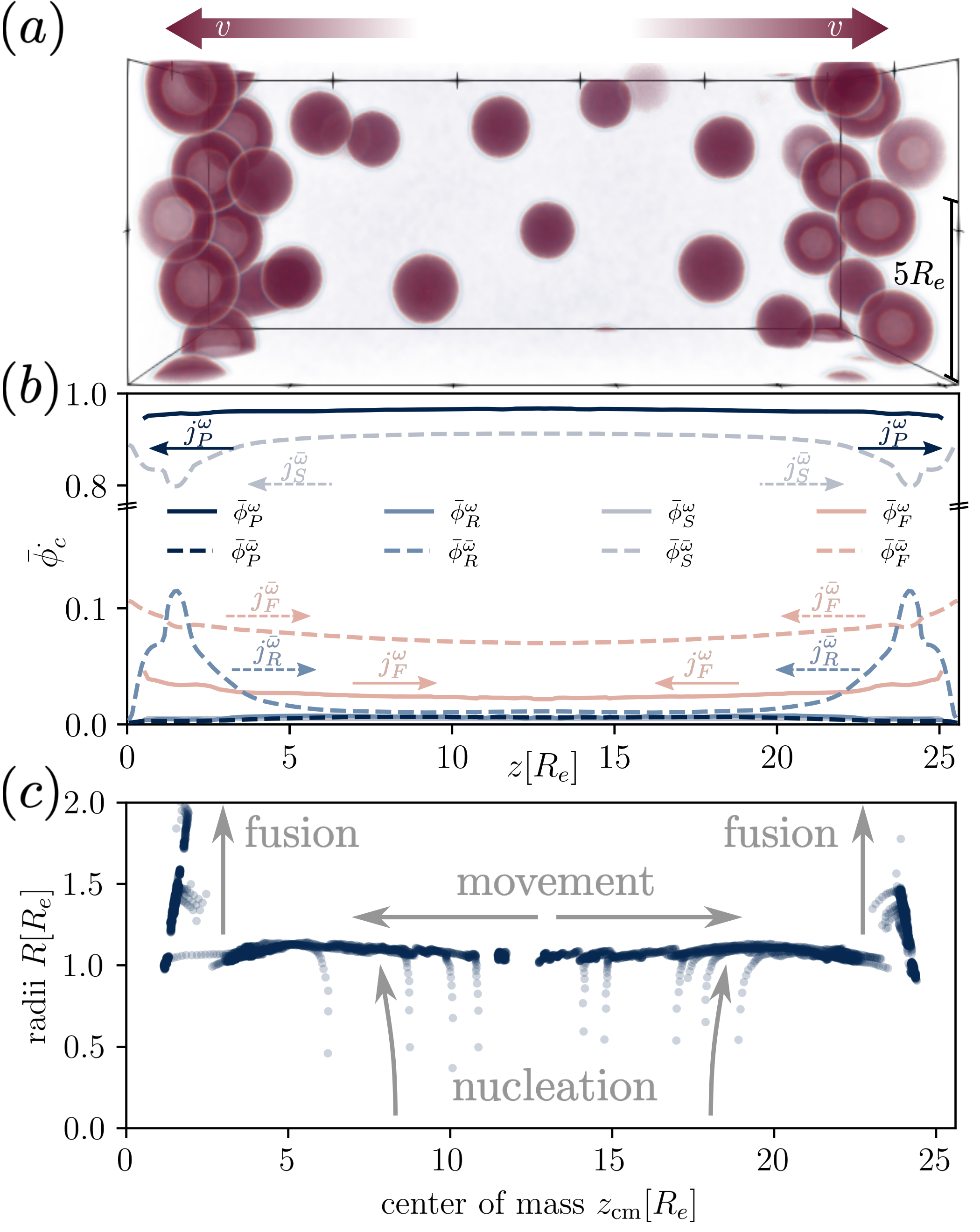}
	\caption{Simulation results for implicit waste treatment  for reaction rate, $r_b=0.08\lambda$. The fuel is nonpreferential towards the product. (a) Example morphology at time $t\lambda=400$ showing the local product concentration $\phi_P(\mathbf{r})$. The movement of droplets is indicated by the arrows. (b) Laterally averaged concentration profiles inside, $\bar\phi_c^{\omega}(z)$ (solid), and outside of the droplets, $\bar\phi_c^{\bar\omega}(z)$ (dashed), as a function of $z$. (c) Individual droplet radii plotted against their center of mass, $z_{\mathrm{cm}}$. The droplet dynamics that determine the shape of the plot are indicated.}
	\label{fig:hp_fg_collage}
\end{figure}

For a backward reaction rate of $r_b=0.08\lambda$ and nonpreferential interaction of fuel and product, $\chi_{PF}N_P=0$, an example morphology is shown in \autoref{fig:hp_fg_collage} (a). Upon start of the simulation, the source injects fuel, $F$, at the sides of the simulation cell and $F$ diffuses into the bulk where the forward reaction commences with precursor molecules, $R$. Since fuel is continuously used within the bulk but only refilled locally, an inhomogeneous but stationary concentration profile builds up quickly, with highest concentration at the source, \ie, $\phi_F$ decreases towards the center of the simulation cell. This implies a continuous flux of fuel towards the center. By virtue of incompressibility, waste, which is part of the solvent and produced in the bulk, flows continuously back. The forward reaction creates product whose concentration quickly surpasses the binodal at $\phi_{P,\mathrm{binodal}}\approx 0.002$ and approaches the mean-field spinodal, $\phi_{P,\mathrm{spinodal}}\approx 0.01$, and phase separates from the solution. The droplet-creation and growth process is illustrated in \autoref{sec:nucleation} of the \ac{SI}.
\nocite{cahn_free_1959}
The coarsening, however, quickly arrests at a finite droplet size that is dictated by the continuous reactions. The hydrophilic precursor has a finite lifetime and thus a finite diffusion length in the solution, and the same holds true for the product in the droplets. Hence, the system locally displays the same behavior as systems with homogeneous fuel concentration. \cite{glotzer_reaction-controlled_1995,zwicker_suppression_2015,hafner_reaction-driven_2023} As before, the nonpreferential interactions between product and fuel lead to a finite fuel concentration inside the droplets. Thus, droplets are transported towards the fuel source, $z=0$ or $L$, leaving large voids at the center of the simulation cell, where the commencement of the forward reaction, again, leads to an oversaturation of product in the aqueous phase, which nucleates new (trailing) droplets. 

Arriving at the sides of the simulation cell, the droplet motion stalls (due to a wall that prevents droplets from screening the fuel source). Here, the droplets fuse with trailing ones, become super-sized (\ie, larger than dictated by the reactions), and immediately shrink again. This fusion-shrinking cycle establishes a statistically stationary state at the boundaries of the simulation cell.

Using a \ac{HKCA}, \cite{hoshen_percolation_1976,hoshen_percolation_1997} we determine the droplet domains, $\omega_i$, where $i$ indexes the droplets and use these to temporally and locally average the concentration profiles inside and outside of the droplet. This is shown in \autoref{fig:hp_fg_collage} (b). Visibly droplets consist of product and fuel only, as the precursor and solvent concentrations practically vanish inside. 

The aqueous phase chiefly consists of solvent, fuel, and precursor. The product density is small in the solvent phase because of the rather large incompatibility between $P$ and $S$, and the resulting small binodal concentration. Nonetheless, the product concentration can locally and temporarily become larger than the averaged concentration when the movement of droplets creates a void space between the droplets at the center of the simulation cell. Subsequently, this gives rise to the creation of a new droplet. 

As the droplets move towards the sides, the product decays into precursor, $R$, and the precursor concentration is large in the aqueous solution near the source. Thus, there is a backward flux of precursor from the sides into the center of the simulation cell that allows to create droplets at the center. In summary, the reaction cycle gives rise to two distinct flux cycles: (i) Fuel is transported from the source into the center, whereas waste flows the opposite way. The (partial) fuel fluxes through the droplets results in a transport of droplets towards the fuel source. (ii) Since product continuously reverts to precursor (reactant), precursor has a high concentration where product droplets move to, \ie, the fuel source. From here precursor flows in opposite direction towards the center. 

The above described life of droplets - creation, movement and fusion - can also be appreciated when using the \ac{HKCA} to plot the droplet radii, $R$, determined as in Ref. \cite{hafner_reaction-driven_2023}, against their center of mass, $z_{\mathrm{cm}}$. This is shown in \autoref{fig:hp_fg_collage} (c). Upon nucleation the droplets quickly approach a uniform size, which varies only slightly over the course of the movement. 
\footnote{A larger simulation-cell size can, indeed, lead to nucleation of small droplets that slowly grow upon approaching the fuel source, as demonstrated in \autoref{sec:choice_of_boundary_conditions} of the \ac{SI}.}
Near the source, however, fusion of droplets leads to droplet sizes that are far larger than the reaction-dictated one. In such cases, the turnover of product on the inside produces more precursor than can diffuse to the outside solution. This leads to an enrichment of precursor, which nucleates a hydrophilic core. This way, a spherical shell emerges. The large shells, in turn, cannot be maintained and shrink. Such vacuolization has recently been shown, both experimentally and theoretically, to be stable in the presence of the here-investigated reaction cycles \cite{bergmann_liquid_2023,bauermann_formation_2023}.
In our case, this happens after rapid fusion, which can also appear at the center of the domain in other scenarios and is not unique to the specific boundary condition.
\footnote{To show that our choice of boundary conditions is irrelevant for the overall effects in the system we performed simulations without the wall, using fully periodic boundary conditions. This leads to fusion of oncoming droplets immediately at the source, $z=0$ aka $z=L$. Here, one also observes the same movement and vacuolization. This is demonstrated in \autoref{sec:choice_of_boundary_conditions} of the \ac{SI}.}

\begin{figure*}
    \centering
    \includegraphics[width=\textwidth]{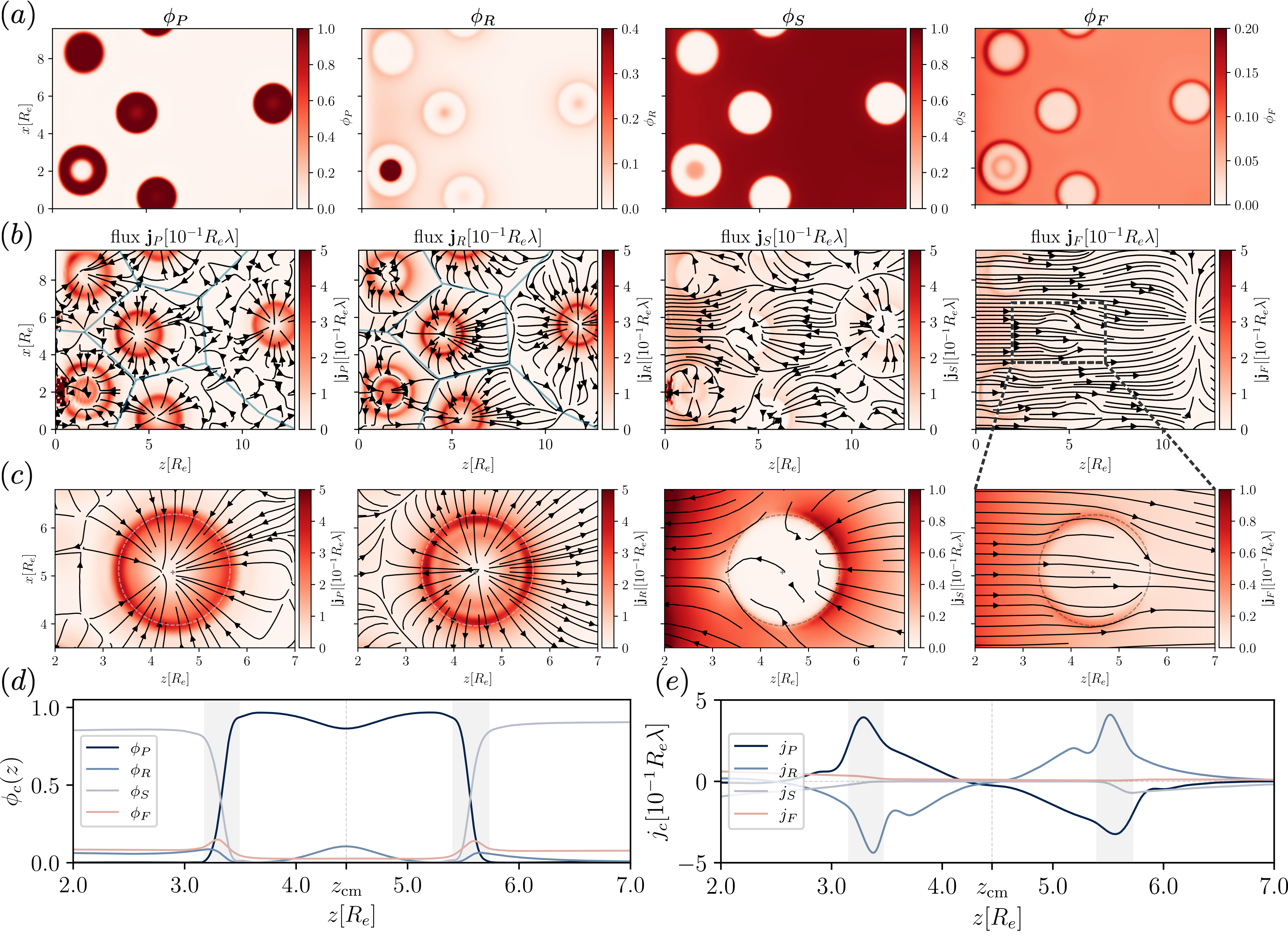}
    \caption{Example concentration fields and fluxes for the simulation shown in \autoref{fig:hp_fg_collage} at time $t\lambda=870$ for for a slice through the domain, showing the $x$-$z$-plane at $y=1.8\R$. (a) Concentration fields, $\phi_c$. (b) Corresponding flux fields, $\mathbf{j}_c$, showing the direction as arrows and the magnitude $|\mathbf{j}_c|$ as the color indicated by the color bars. The Voronoi tessellation is indicated in turquoise. (c) Flux fields, zoomed into a single droplet, as indicated for $\mathbf{j}_F$. The center of mass and interface position are indicated. (d) Concentration profiles for a line at $x_{\mathrm{cm}}=5.05\R$ and (e) the corresponding fluxes $j_c:=\mathbf{j}_c\cdot \hat e_z$. In the latter two panels, the droplet center and the interface region are indicated, as dashed vertical line and shaded region, respectively.}
    \label{fig:hp_fg_fluxes2d}
\end{figure*}
The directed movement of droplets participating in the reaction cycle is also rooted in the presence of concentration gradients, \ie, \textit{reaction-driven diffusiophoresis}. It is similar to the case of passive droplets but is additionally influenced by the reactions. Since the fluxes in this system can become rather complex, let us have a detailed look at these for a representative snapshot, plotting a \ac{2D} slice through the $x$-$z$-plane of the \ac{3D} domain. \autoref{fig:hp_fg_fluxes2d} shows the concentration fields, the resulting fluxes, and a zoom-in on a single droplet in an example morphology (\ac{2D} slices and corresponding profiles through the center of the droplets). For this, we filtered the thermal noise by running the simulation without it for a short time, $t\lambda=0.1$, as described in \autoref{sec:analysis_procedure} in the \ac{SI}.

We can see that the droplets consist majorly of product, with a small fuel concentration remaining, which is slightly enriched at the interface to screen the interactions of hydrophobic product and solvent. The precursor concentration, $R$, is high at the center of the droplet and decreases close to the interface towards equilibrium concentration. In the case of a spherical shell the precursor is strongly enriched on the inside, while the opposite is true for fuel and solvent.

From the flux plots we can see that there is a continuous eflux of precursor, $R$ out of the droplets and a concomitant influx of product. It is largest close to the interface. This is the result of the driven chemical reactions: On the one hand, the backward reaction produces hydrophilic precursor in the hydrophobic environment of a droplet, which diffuses out of the droplet. On the other hand, the forward reaction creates hydrophobic product in the aqueous solution by conversion of the high-energy fuel molecules. This leads to a product flux towards the droplets on the outside. The behavior is studied in detail in Ref.~\cite{zwicker_growth_2017}.

In our scenario, droplets are usually densely packed and therefore compete for product. In \autoref{fig:hp_fg_fluxes2d} (b) we can observe that product fluxes always point into the direction of the nearest \ac{RDA}-formed droplet. Using a Voronoi tessellation. \cite{aurenhammer_voronoi_2000,okabe_spatial_2009} we assign a region surrounding each droplet as the accumulation volume. These  Voronoi domains, $\Omega_i$, with $i$ indexing the droplets are delineated by a turquoise line in the plot. The morphology of these domains is analysed in detail in \autoref{sec:reaction_contributions} in the SI, for representative scenarios, and we discuss their influence on droplet movement. 

From the close-ups of the flux fields we can see that the solvent flux almost vanishes inside the droplet, whereas the fuel, though slightly deflected, flows through it. This is similar to the case of passive droplets in an external gradient. Product and precursor gain additional radial fluxes, towards and away from the center of the droplet, respectively. While the precursor flux appears mirror symmetric with respect to the droplet's center, the product flux is slightly off-centered. This can also be appreciated in the profile of \autoref{fig:hp_fg_fluxes2d} (e): The product flux is slightly smaller than zero in the droplet center, whereas the fuel flux is slightly positive inside the droplet. 
 
From \autoref{fig:hp_fg_fluxes2d}, we can qualitatively identify three contributions that lead to movement of the \ac{RDA}-formed droplets: (i) The finite fuel flux inside the droplets demands a product flux\footnote{In principle the flux can also be met with a precursor flux, but while the precursor flux can be high inside, the concentration is low and from \autoref{eq:flux-lagrange-multiplicator-eliminated} we see that it will not gain a noticeable contribution.} in opposite direction\footnote{This is also the case with reaction, because, in our case, these conserve volume.}. In principle, the same applies for the waste flux, which in this scenario vanishes. (ii) The production inside the accumulation volume creates product on the outside of the droplet. Asymmetry in production can thus lead to movement, if the flux is higher on one side. Such asymmetry can arise both, from concentration gradients of fuel and precursor or from differences in accumulation volume in front and behind the droplet, respectively. 
\footnote{\ac{RDA}-formed droplets are continuously resupplied with product that originates in its surroundings. Here, we assume the droplets are exclusively supplied by product generated in its Voronoi cell, $\Omega_i$. In the absence of concentration gradients, a droplet adopts a finite size, for which the loss of droplet material (product), which reverts into precursor inside the droplet, matches the gain due to the forward reaction, $R+F \to P+S$, in its Voronoi cell. In our scenario, however, precursor and fuel have a higher concentration closer to the fuel source. Hence there is an imbalance in production that provides an additional cause of directed droplet movement towards the fuel source.} (iii) Since the product density inside the droplet is higher away from the fuel source, the backward reaction may also lead to asymmetric flows of precursor and hence of product, which would shift the droplet's positions.

In \autoref{sec:velocities_analytical} of the \ac{SI}, we analytically calculate the velocity of \ac{RDA}-formed droplets, thereby quantifying the above mentioned contributions. To this end, we use the approximations that (i) interfaces are sharp, (ii) product material that is generated inside the Voronoi cell, $\Omega_i$ of the individual droplet, $i$, instantly condensates on the surface of the droplets, and (iii) product that decays inside the droplet is instantly refilled from the interface. The latter two approximations, the \textit{instantaneous condensation approximations} (ICA), essentially replace the fluxes that arise from forward and backward reaction of product in \autoref{fig:hp_fg_fluxes2d} by sources and sinks immediately at the droplet interface and retain only the (nonreactive) fluxes that arise from fuel and waste through incompressibility, \ie, passive diffusiophoresis of liquid condensates, as discussed in \autoref{sec:passive_diffusiophoresis}. 
The approximation is valid if the concentration fields change only slowly during the diffusion time of a single molecule across the accumulation volume. 

With these approximations we can now identify the (nonreactive) product flux inside the $i^{\text{th}}$ droplet, $\mathbf{r}\in \omega_i$, as in the case of passive diffusiophoresis, $\mathbf{j}^{\text{nr}}_P(\mathbf{r}) = -[ \mathbf{j}_F(\mathbf{r}) + \mathbf{j}_W(\mathbf{r})]$. We obtain for the droplet's velocity, $v_i$ by
\begin{align}
		 v_i \int_{\omega_i}\hspace*{-0.2cm}\mathrm{d}\mathbf{r}\,&\phi_P(\mathbf{r}) = - \int_{\omega_i}\hspace*{-0.2cm}\mathrm{d}\mathbf{r}\,\left[j_F(\mathbf{r})+ j_W(\mathbf{r})\right] \nonumber \\
   &+ R \int_{\Omega_i}\hspace*{-0.2cm} \mathrm{d}\mathbf{r}\, [r_f\phi_R\phi_F-r_b\phi_P] \zeta(|\mathbf{r}-\mathbf{r}_{\rm cm}|)\cos\theta\label{eq:velocity_symmetric_final}	
\end{align}
where $j_c:=\mathbf{j}_c\cdot\hat{\mathbf{e}}_z$ are the longitudinal components of the fluxes. The factor $\zeta(|\mathbf{r}-\mathbf{r}_{\rm cm}|)\cos \theta$, with $\theta$ being the angle of $\mathbf{r}-\mathbf{r}_{\rm cm}$ to the transport direction, $\hat{\mathbf{e}}_z$, accounts for both, the finite lifetime of the product during which it reaches the droplet surface and the location on the droplet surface, at which the product attaches after it has been generated inside the Voronoi cell. 

\autoref{eq:velocity_symmetric_final} identifies the three different contributions to the droplet's velocity: (i) the first integral on the r.h.s.~denoted as \textit{nonreactive flux contribution}, (ii) the \textit{forward reaction contribution}, corresponding to the first summand in the second integral, and (iii) the \textit{backward reaction contribution}, represented by the second term of the second integral.

\begin{figure}
    \centering
    \includegraphics[width=0.48\textwidth]{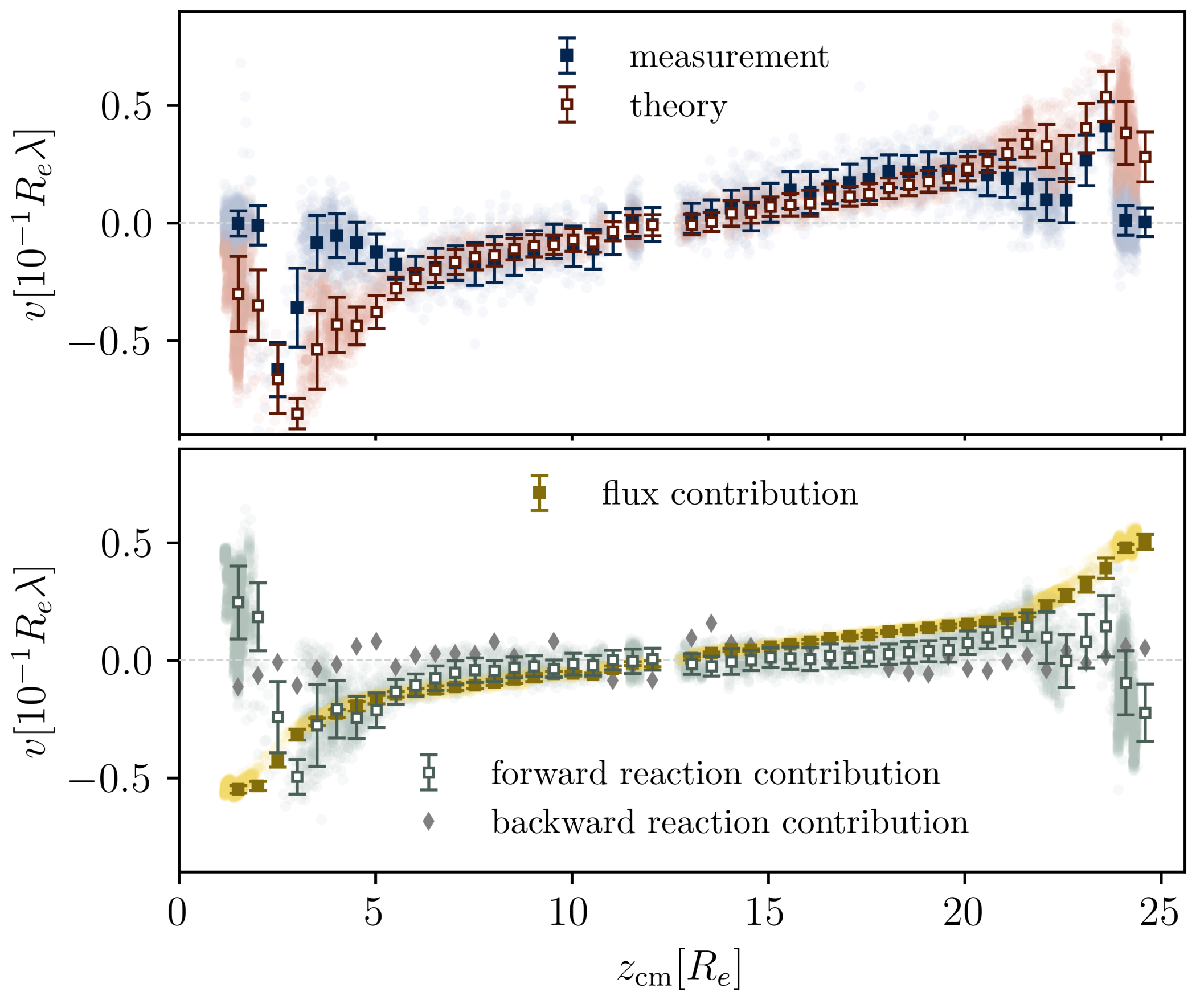}
    \caption{Velocities in vertical direction, $v$, and their contributions plotted against the droplet centers, $z_{\mathrm{cm}}$. (a) Mean measurement of droplet velocities using the \ac{HKCA} (blue) and mean prediction of \autoref{eq:velocity_symmetric_final} (red). The individual results are plotted as dots in light colors underneath and the error bars indicate the spread among these. (b) Decomposition of the prediction into the three contributions, flux contribution (yellow), forward reaction contribution (turquoise), as well as the backward reaction contribution (grey). For the former two we show individual droplet predictions in light colored dots, but omit this for the last one for clarity of the figure, where individual droplet contributions are uniformly spread around zero.}
    \label{fig:hp_fg_velocities}
\end{figure}

An analytical prediction is not available  but we can use the measured concentration profiles and concomitant fluxes, droplet radii, and Voronoi cells to compare the prediction of \autoref{eq:velocity_symmetric_final} with the simulation data for the droplet velocity, $v$. \autoref{fig:hp_fg_velocities} shows the measured droplet velocities, obtained by averaging over velocities of all droplets at position, $z=z_{\rm cm}$, and the theoretical prediction. The prediction matches the measurement in the bulk but deviates close to the source, where the prediction overestimates the velocity. From \autoref{fig:hp_fg_collage} (a) we infer that the deviation occurs where droplets collide with the layer of shells and fuse -- phenomena that are not consider in \autoref{eq:velocity_symmetric_final}.

The lower panel of \autoref{fig:hp_fg_velocities} breaks down the prediction into its three contributions. The nonreactive flux contribution is the dominant driving force for movement towards the source. Inside the bulk, $5 \leq z/\R \leq 20.6$, the forward reaction contribution is slightly smaller and acts in the same direction. Close to the source, however, it becomes small and can act even in opposite direction, due to highly asymmetric accumulation volumes, $\Omega_i$, as described in detail in \autoref{sec:reaction_contributions} of the \ac{SI}. The backward reaction contribution is largely negligible.

\subsubsection{Implicit Fuel} 
\label{sec:implicit_fuel}

\begin{figure}
	\centering
	\includegraphics[width=0.48\textwidth]{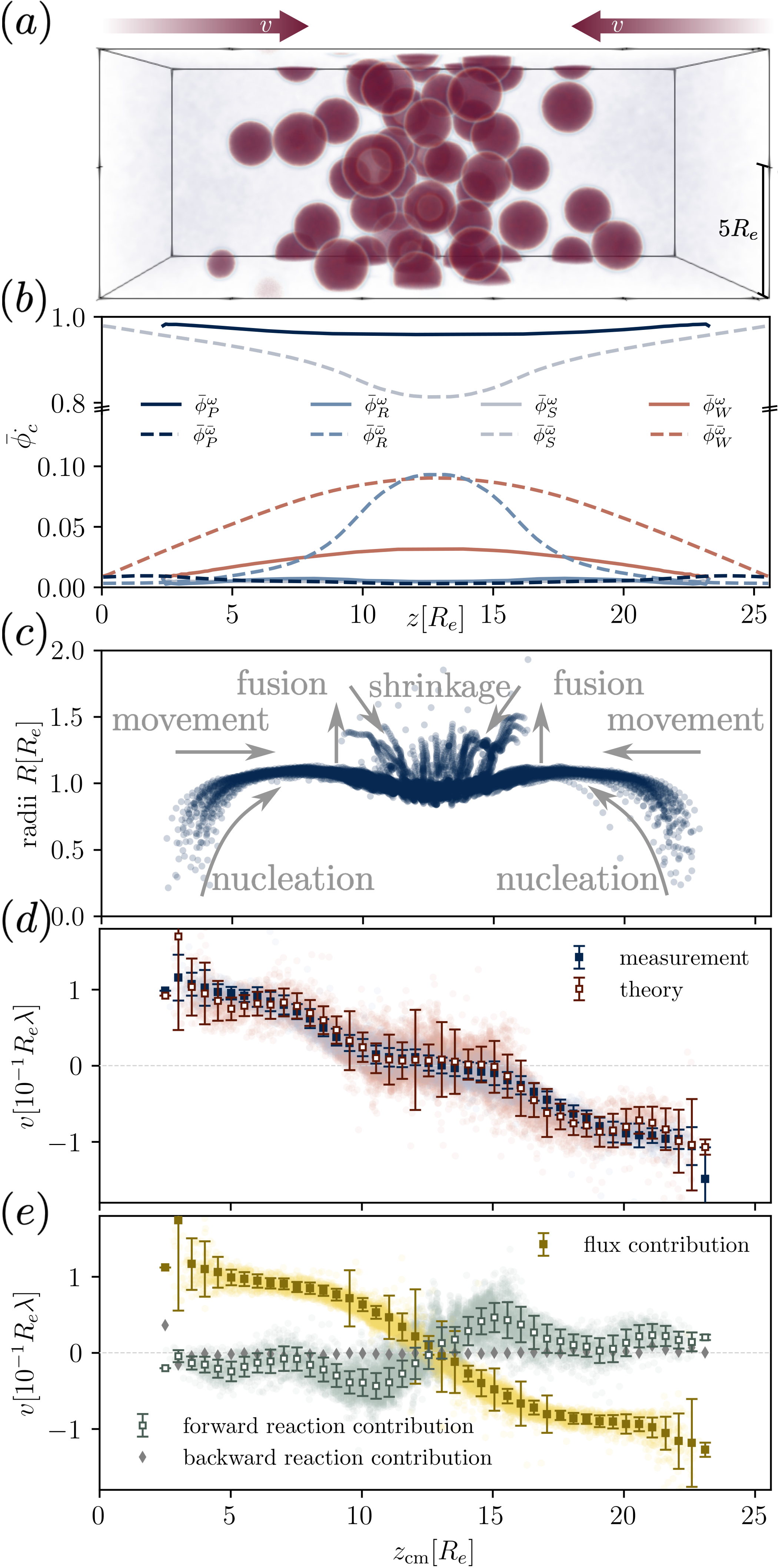}
	\caption{Simulation results for implicit treatment of fuel for reaction rate constant $r_b=0.08\lambda$ and neutral interaction $\chi_{PW}=0$. (a) Example Morphology at $t\lambda=1000$. The movement of condensates is indicated. (b) Laterally averaged concentration profiles inside, $\bar\phi_c^{\omega}$ (solid), and outside the droplets, $\bar\phi_c^{\omega}$ (dashed). (c) Droplet radii, $R$, as obtained from the \ac{HKCA}. The dynamics seen in the simulations are indicated. (d) Measurements and predictions of vertical droplet velocity, $v$, plotted against the center of mass, $z_{\mathrm{cm}}$. (e) The three contributions of the prediction from \autoref{eq:velocity_symmetric_final} plotted against center of mass, $z_{\mathrm{cm}}$.}
	\label{fig:wg_velocity_collage}
\end{figure}
In the previous section waste was treated as part of the solvent, \ie, strongly repulsive towards the product. Importantly, since the fuel source also acts as a waste (solvent) sink in this setup, the waste that is produced in the bulk diffuses towards the sink. This implies a waste-flux effect that is opposite to the one of fuel flux if there are more favorable interactions of waste towards the product. To show this, we explicitly treat waste with $\chi_{WP}N_P=0$ but fuel implicitly, as part of the solvent. The forward reaction is then described by $R + S \rightarrow P +W$ and commences with a rate constant, that is ten times higher than before, $r_f = 4 r_b$, in order to to keep the conversion rate comparable to the previous set-up.
The simulation results are shown in \autoref{fig:wg_velocity_collage}. In contrast to the previous set-up, there is a directed movement of droplets to the center of the domain. As droplets move towards the center collectively, large voids arise, in which the fuel commences to react with precursor. Product becomes supersaturated in this part of the solution and, eventually, a spherical droplet nucleates, which starts moving. In this scenario, droplets grow rapidly after nucleation and only become a little smaller before the finite size of the simulation cell gives rise to collision with oncoming droplets from the other side, \autoref{fig:wg_velocity_collage} (a). This dynamics can also be identified in the individual droplet radii as a function of their center of mass, presented in \autoref{fig:wg_velocity_collage} (c).  At the center, fusion events result in droplets that become too large to be maintained externally by \ac{RDA} and vacuolize, see \autoref{fig:wg_velocity_collage} (a). 
In the laterally averaged concentration profiles, shown in \autoref{fig:wg_velocity_collage} (b), one can see that, following the product droplets' movement into the center, the precursor is enriched in the solution at the center. Waste also has the highest concentration in the solution at the center, whereas it is slightly lower inside the droplets, for entropic reasons. The inhomogeneous waste profile implies a flux of waste towards the sink on the boundaries, $z=0$ and $L$. Finally, the local velocities are measured and compared to the theoretical prediction in \autoref{fig:wg_velocity_collage} (d), where the prediction matches the measurements precisely. 
\autoref{fig:wg_velocity_collage} (e) dissects the prediction into its contributions: The nonreactive flux contribution stems from the waste flux from the center to the sink and its coupling to the product flux \textit{via} incompressibility. This contribution sets the overall nature of the droplet movement. The forward reaction contribution is large close to the center and is opposed to the nonreactive flux contribution. It gives rise to the 'shoulder' that is visible in the velocity profile at the center. The shape of the contribution is dictated by the strong asymmetry of accumulation volumes, where droplets have a significant larger volume behind than in front, when colliding with the crowd of droplets at the center. Whereas the former contribution is largely independent of the droplet arrangement within the simulation cell, this contribution depends heavily on the arrangement and overall movement of droplets. Hence, the nature of the forward reaction contribution is determined by the nonreactive flux contribution. Again, the backward reaction contribution fluctuates around zero and is negligible. 

\subsection{RDA-formed Droplets with Explicit Fuel and Waste}
\label{sec:explicit-fuel-waste}

The previous two scenarios showcased two diametrically opposed effects, movement of \ac{RDA} droplets towards the fuel source, mainly due to a diffusive flux of fuel from the source into the bulk, and movement of these away from the waste sink, due to a diffusive flux of waste from the bulk to the sink. Treating now both components explicitly, with neutral or favorable interactions towards the product, and locating fuel source and waste sink in the same place will result in both effects working against each other. Indeed, if the two components have the same properties, \ie, interactions and mobilities, both flux contributions cancel exactly. We detail this argument in \autoref{sec:symmetric_interactions} of the \ac{SI} and show supporting simulations. After an initial transient, when the fuel flux dominates because waste is not yet present, the movement of droplets towards the sides slows down and, eventually, completely halts.

\begin{figure*}
	\centering
	\includegraphics[width=\textwidth]{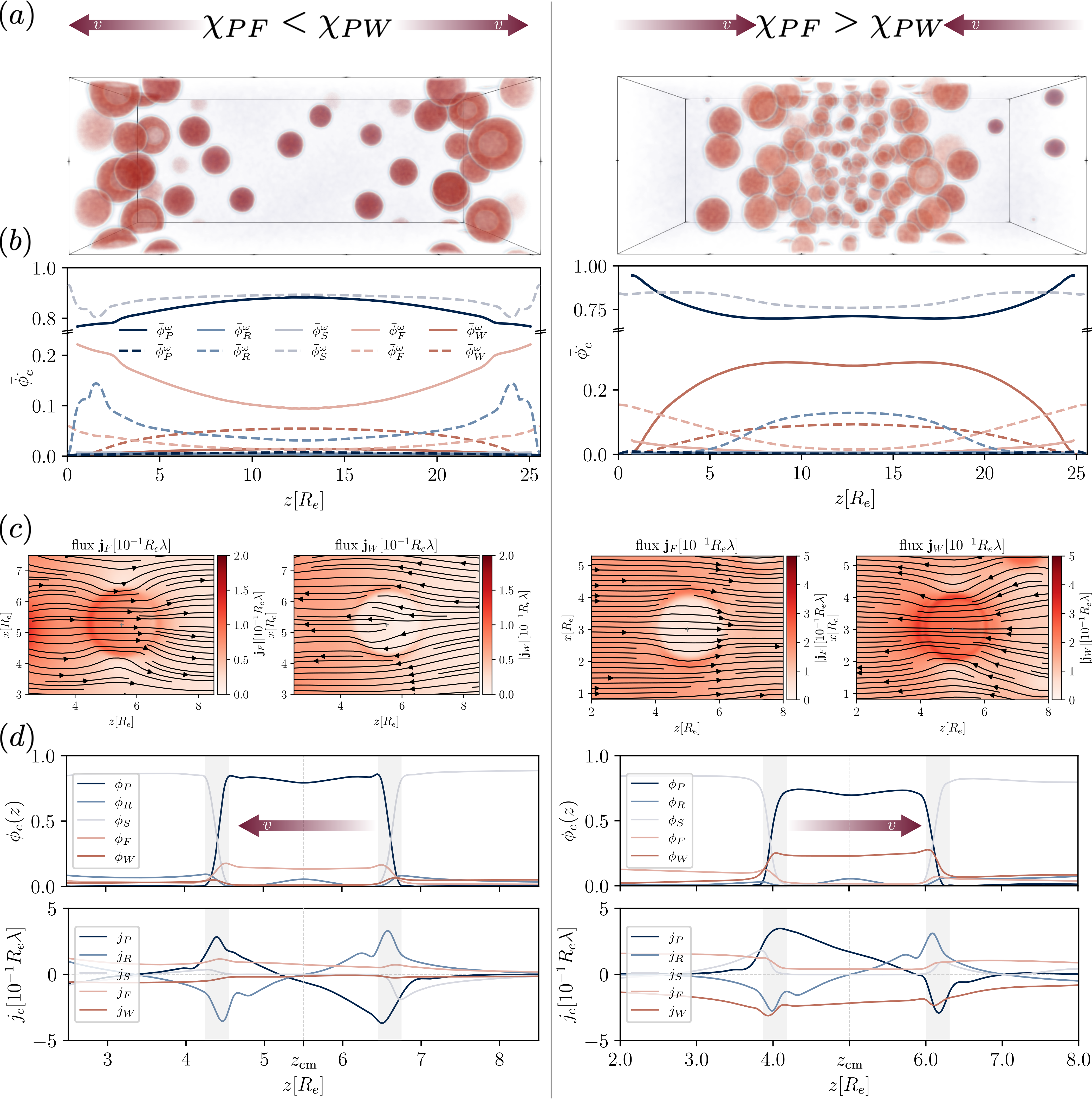}
	\caption{(a) Example morphologies for two interaction scenarios for reaction rate, $r_b=0.08\lambda$. On the left, fuel has a higher affinity towards product than waste ($\chi_{PF}N_P=-40,\,\chi_{PW}N_P=0$), leading to movement of product droplets towards the fuel source. On the right, waste has the higher affinity ($\chi_{PF}N_P=0,\,\chi_{PW}N_P=-40$), leading to movement towards the center. (b) Temporally and laterally averaged concentration profiles, inside the droplets, $\bar\phi_c^{\omega}$ (solid), and outside, $\bar\phi_c^{\bar\omega}$ (dashed). (c) (Noise-smoothed) Flux field of fuel, $\mathbf{j}_F(\mathbf{r})$, and waste, $\mathbf{j}_W(\mathbf{r})$, showing a droplet on the left side of the domain. (d) Concentration profiles (top) and flux profiles (bottom) through the center of the droplet of panel (c) along the $z$-direction. The direction of transport is indicated. The droplet interface is indicated in gray.}
	\label{fig:fwg_morphologies_and_profiles}
\end{figure*}

In natural systems, this symmetry between fuel and waste with respect to their interactions and mobility can be broken, allowing for an exquisite control over the directed motion of droplets. Therefore, let us consider different scenarios where either the fuel or waste has an affinity for the product, $\chi_{PF}N_P<0$ or $\chi_{PW}N_P<0$, whereas the other component interacts neutrally. The attractive component is enriched inside the product droplets which results in its flux contribution to dominate and, once again, we observe directed movement of the \ac{RDA} droplets. Specifically, if $\chi_{PF}<\chi_{PW}$, droplets move towards fuel source and waste sink. On the contrary, if $\chi_{PF}>\chi_{PW}$ droplets move away from it, towards the center of the domain. This is demonstrated in \autoref{fig:fwg_morphologies_and_profiles} (a). Example morphologies are given for these two scenarios. 

\autoref{fig:fwg_morphologies_and_profiles} (b) shows temporally and laterally averaged concentration profiles, outside, $\bar\phi_c^{\bar\omega}$, and inside, $\bar\phi_c^{\omega}$, of the droplets for the two different scenarios. Visibly, the component which has an affinity towards the product is significantly enriched inside the droplets, whereas the nonpreferential component is depleted. Consequently, the attractive component preferentially flows through the droplets, determining their directed motion. \autoref{fig:fwg_morphologies_and_profiles} (c) demonstrates this behavior for the example of a single droplet for the two scenarios. Finally, \autoref{fig:fwg_morphologies_and_profiles} (d) shows the concentration and flux profiles for a cut through the center of the example droplets of panel (c). The direction of movement is indicated by the arrows. Again, one can observe the enrichment or depletion of fuel or waste, as well as the slight enrichment of precursor in the droplet center. From the flux profiles, we can identify once more the three contributions: Driven by the forward reaction, there is an influx of product on the outside, potentially asymmetric. Driven by the backward reaction, precursor flows from the inside towards the outside, approximately spherically symmetric. Fuel and waste both flow through the droplet and their combined flux determines the backward flux of product in the droplet center. Visibly, there is $j_P(z_{\mathrm{cm}}) \approx j_P^{\text{nr}}(z_{\mathrm{cm}}) \stackrel{\mathrm{def.}}{=} -[j_F(z_{\mathrm{cm}})+j_W(z_{\mathrm{cm}})]$ in both scenarios, leading to the corresponding droplet motion.

\begin{figure}
    \centering
    \includegraphics[width=0.48\textwidth]{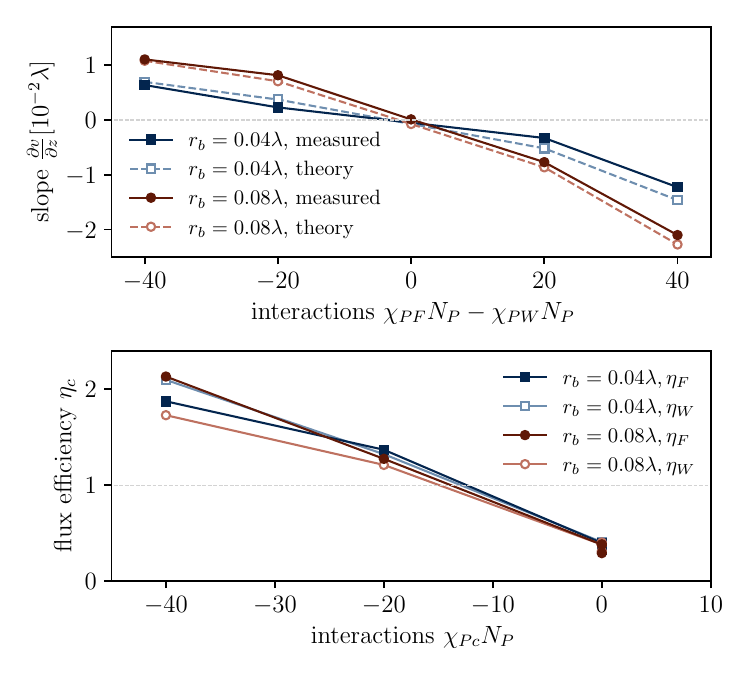}
    \caption{Top: Indicator for the droplet dynamics, the slope, $\partial v/\partial z$, of a linear fit through the velocity profiles in the bulk plotted against the interaction difference $(\chi_{PF}-\chi_{PW})N_P$, where one of the interactions always vanishes and the other one is negative. Thus, for negative values fuel has an attraction towards the product and for positive ones waste is attractive. Bottom: Efficiency measurement, $\eta_c$ of the components $c=F,W$ for all systems. }
    \label{fig:fwg_velocity_slopes}
\end{figure}
The simulation results for a variation of interactions and reaction rate are depicted in full detail in \autoref{sec:simulation-results-explicit}, where we show the radii, velocity measurement and theoretical comparison, as well as the individual contributions to the velocity for each parameter. We introduce a crude estimator for the droplet dynamics, by taking the slope of a linear fit through the velocity measurements in the bulk of the domain, $5\leq z\leq 21.6\R$. The linear fit is depicted in \autoref{fig:fwg_velocity_vs_position} (b) of the \ac{SI}.
Its slope adopts positive values for movement towards the fuel source, while it becomes negative for droplet movement away from the source into the bulk. We compare the slope for the measurement and the theoretical estimation of \autoref{eq:velocity_symmetric_final} in \autoref{fig:fwg_velocity_slopes} (top). Clearly, the more attractive one of the components becomes, the faster the droplet movement becomes  towards its source. A higher reaction rate also leads to faster movement, since concentration gradients become stronger. Moreover there is an asymmetry in the plot, indicating that we observe higher velocities in the case of waste-attractive scenarios than in the fuel-attractive ones.

Additionally we quantify the flux efficiency through the droplet by comparing the fuel and waste flux through the droplet to the laterally averaged flux (corresponding to the uniform flux of the system without droplets). This motivates us to define the efficiency, $\eta_c = \frac{1}{L_z}\int \mathrm{d}z\, \bar j_c^\omega(z)/\bar j_c(z)$, for $c=F,W$.  \autoref{fig:fwg_velocity_slopes} (bottom) shows the efficiency plotted against the respective interaction parameter. Clearly the attractive component, which is enriched inside the droplets experiences are significant efficiency increase. For the strongest attractions that we have probed, the flux increases by a factor of $\eta_c\approx 2$ on the inside compared to the outside, while it is decreased to $\eta_c\approx 0.3$ for nonpreferential interactions of fuel or waste.

To corroborate the continuum model results, we perform particle-based simulations of a soft, coarse-grained model for polymer melts, making use of the single-chain in mean field approximation \cite{daoulas_single_2006,schneider_multi-architecture_2019}.
\nocite{dreyer_simulation_2022}
In \autoref{sec:particle-based} of the \ac{SI} we show that we observe the same droplet dynamics and interaction dependencies.

\paragraph*{Diffusiophoresis Velocity from Metabolism}

Estimating the significance of the effect in living cells, we take the metabolite ATP as an example. Its molecular concentration gradient, as estimated by Sear in Ref. \cite{sear_diffusiophoresis_2019} can become roughly $|\nabla \rho_{\mathrm{ATP}}|\sim 10^5-10^6/\mu \mathrm{m}^{-4}$ for  molecular concentrations on the order of $\rho_\mathrm{ATP}\sim 10^7\mu \mathrm{m}^{-3}$. Taking a molecular volume of $1/\rho_0\sim 10^{-9} \mu \mathrm{m}^{3}$ results in a normalized volume concentration of $\phi_\mathrm{ATP}=\rho_{\mathrm{ATP}}/\rho_0=10^{-2}$ and a gradient of $|\nabla \phi_\mathrm{ATP}|=|\nabla \rho_\mathrm{ATP}/\rho_0| \sim 10^{-4}-10^{-3} \mu \mathrm{m}^{-1}$. With the diffusion coefficient of $D_{\mathrm{ATP}}\sim 10^2 \mu \mathrm{m}^2/\mathrm{s}$ \cite{de_graaf_vivo_2000} this results in an ATP flux of $j_\mathrm{ATP} \approx D_\mathrm{ATP} |\nabla \phi_\mathrm{ATP}| \sim 10-100 \mathrm{nm}/s$ inside the solution. Hence, this is the order of magnitude, one can expect for the transport velocity at neutral interactions, independent of condensate size. This constitutes just one component, but in cells concentration gradients can be present in an abundance of components. With attractive interactions, the transport efficiency can even be increased. 

\subsection{Multiple Droplet Types}
\label{sec:multiple-droplet-types}

\begin{figure}
    \centering
    \includegraphics[width=0.48\textwidth]{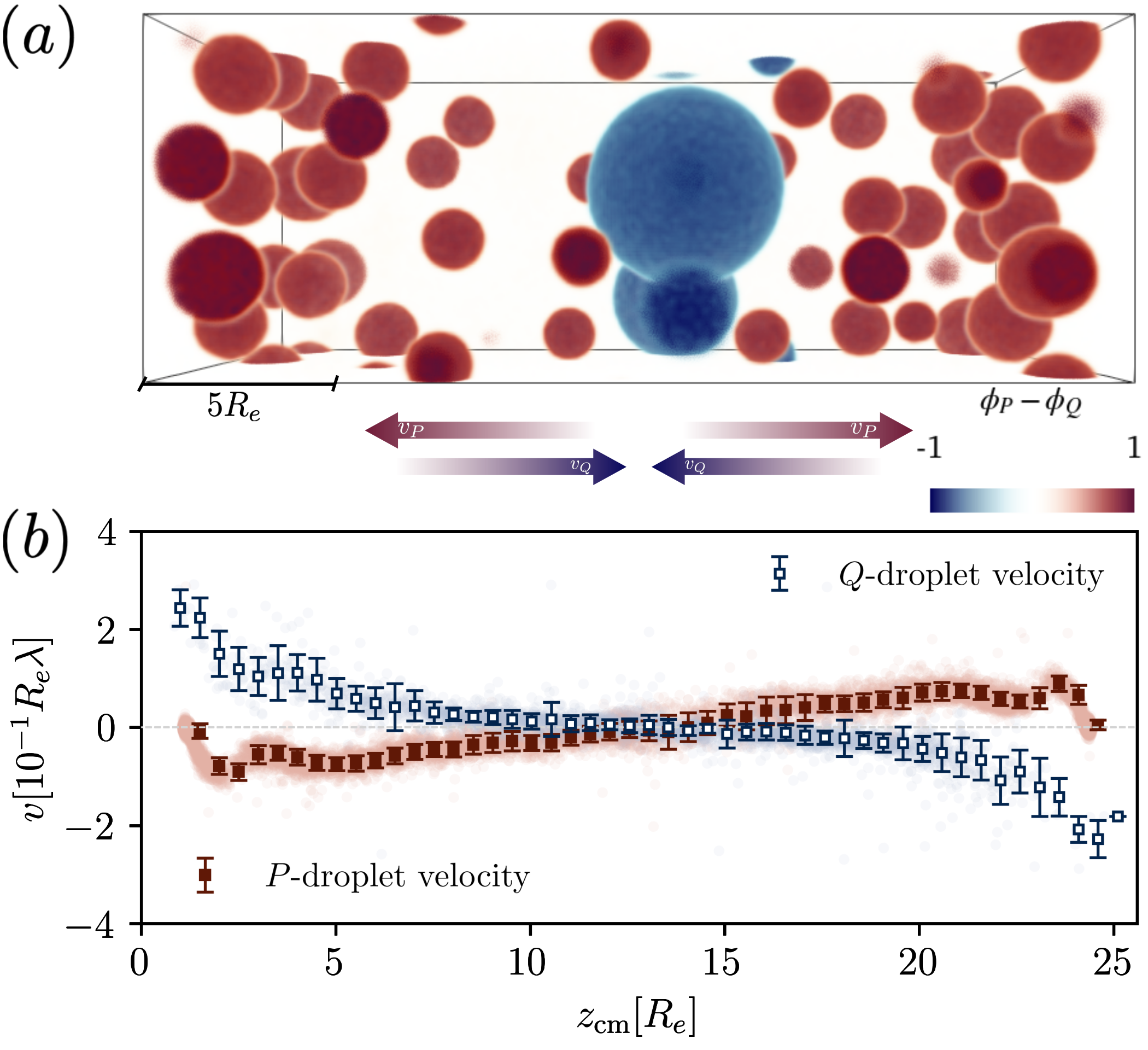}
    \caption{Simulations with additional waste-attractive component $Q$, which is not involved in the reaction cycle. (a) Example snapshot after the $Q$-rich droplets have moved to the center $t\lambda=500$. (b) Velocity measurements of $P$- (red) and $Q$-rich (blue) droplets plotted against droplet position.}
    \label{fig:multidroplets}
\end{figure}
We note, that in biological systems, there typically exists a plethora of phase separating quantities, \cite{uversky_intrinsically_2017,zwicker_evolved_2022,jacobs_theory_2023,thewes_composition_2023} inducing the formation of many moieties with entirely different composition. Here we demonstrate that the existence of a reaction cycle can lead to movement of all of such droplets, whether these are involved in the reaction or not. All of this only depends on the interactions with fuel and waste. This can lead to a subset of droplets being driven towards the center, \ie, the moieties that interacts favorably with the waste, whereas another subset is driven towards the fuel source and waste sink, \ie, the ones that interact favorably with the fuel. We show the phenomenon for two types of droplets with different compositions and interactions with fuel and waste. To this end, we introduce another macromolecular species, $Q$, that phase separates from solution but is not involved in the reaction cycle. Introducing an incompatibility between the two hydrophobic liquids, $P$ and $Q$, results in the formation of two distinct droplet types. The parameters and simulation protocol is described in \autoref{sec:multiple-droplet-types} of the \ac{SI}. Interacting favorably with the waste, the $Q$-rich droplets move towards the center of the domain and coalesce, while the $P$-rich droplets that are formed by \ac{RDA} continue to nucleate in the center and move towards the source at the boundary of the simulation cell. The final $Q$-rich droplet is thus driven into a non-equilibrium state in which the diffusion away from the center is hindered such that it stays positioned in the center. 
\autoref{fig:multidroplets} (a) shows an example snapshot after all of the (passive) $Q$-rich droplets have been transported to the center, while the $P$-rich droplets stay dynamic without reaching a stationary state. \autoref{fig:multidroplets} (b) shows the corresponding velocity measurements for the two droplet types, where the opposed movement becomes clear. Hence, this is a reliable but very primitive method for (proto-) cells, to organize their inside components.

\subsection{Growing complex reaction-driven assemblies}
\label{sec:complex-assemblies}
In the previous scenarios, we considered a simple phase-separating product, whose droplets did not exhibit internal structure and no free-energy barriers existed that could hinder merging of aggregates. Here, we will consider a reaction cycle, in which the product is amphiphilic, instead of simply hydrophobic. Such a system has been introduced in Ref.~\cite{hafner_reaction-driven_2023}, illustrating the general, universal aspects of reaction-driven amphiphilic self-assembly. We model the amphiphilic product as a diblock copolymer, with a hydrophilic $A$-head and a hydrophobic $B$-tail. To do so, we employ the free-energy derived by Uneyama and Doi, \cite{uneyama_density_2005,uneyama_calculation_2005,uneyama_density_2007} which reduces to the previous free-energy for single-block molecules and to the Ohta-Kawasaki model \cite{ohta_equilibrium_1986} close to the onset of micro-phase separation. The continuum model allows to blend different molecular architectures, which is required here.

We provide a description of the simulations and parameter details in \autoref{sec:parameters_amphiphiles} of the \ac{SI}. The amphiphilic product aggregates into micelles or vesicles. Most importantly the hydrophobic tails of the product are repulsive towards precursor, solvent, fuel, and waste, \ie, there is no nonreactive flow across the hydrophobic aggregate core. However, the hydrophilic head group interact favorably with the fuel.
In the absence of a concentration gradient, the stable morphology consists of stacked planar bilayers (lamellar phase) or vesicles. Forming such aggregates from an initially homogeneous solution, however, requires topological changes from micelles to vesicles or merging of aggregates that are hampered by a free-energy barrier. This distinguishes the behavior from self-assembled, amphiphilic aggregates from droplets that readily merge. Thus, in the absence of motion, amphiphilic systems are likely to become trapped in intermediate morphologies such as micelles (see \autoref{fig:amphiphile-stationary-test} of the \ac{SI}). 

If, however, fuel is locally refilled, the concomitant nonreactive fuel flux from the sides of the simulation cell towards the center will generate an opposite flux of aggregates with fuel-attractive head group towards the boundaries of the simulation cell. This motion allows for a slow growth of the aggregates as it moves towards higher fuel concentration.

\begin{figure}
    \centering
    \includegraphics[width=0.48\textwidth]{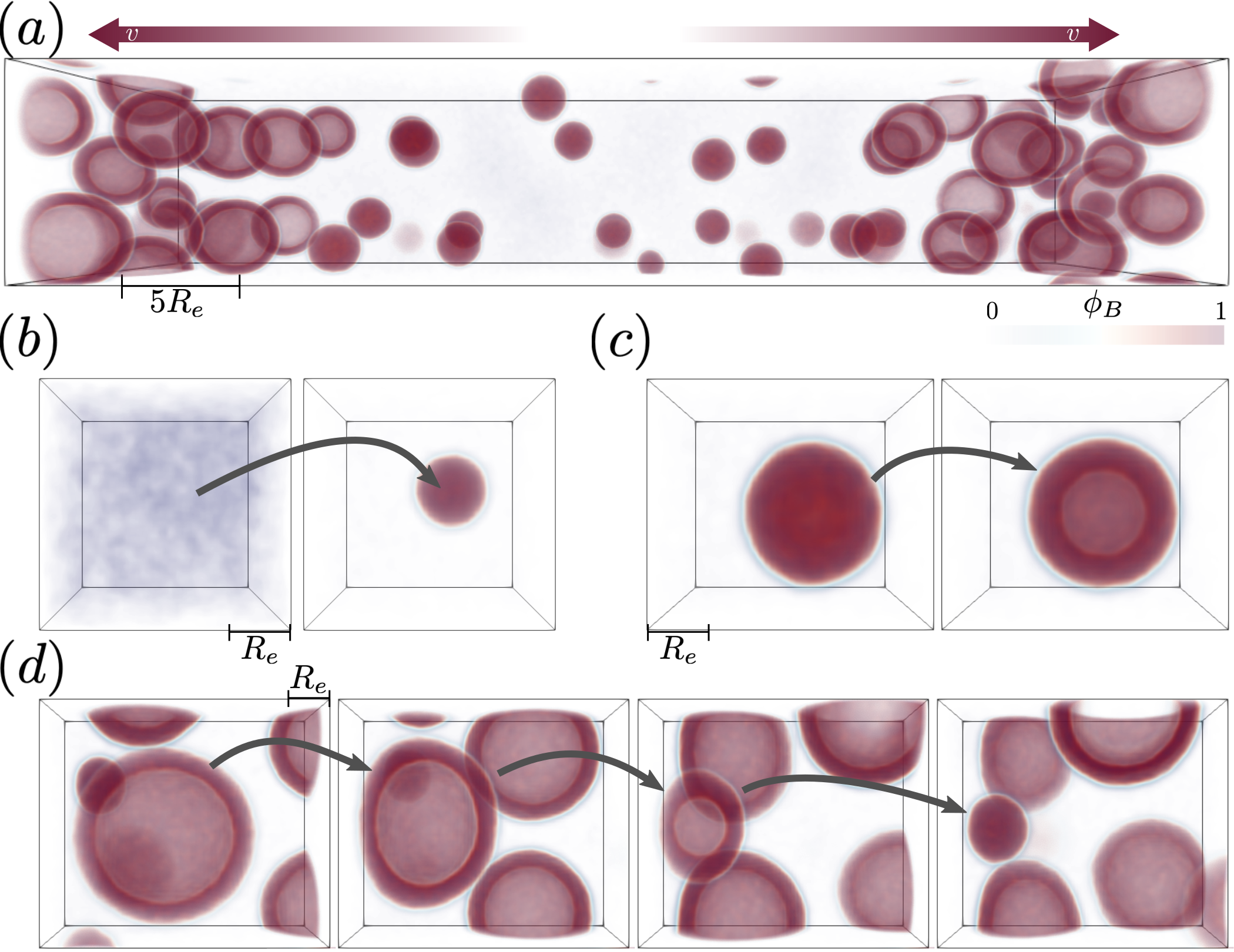}
    \caption{Example snapshot of a system with amphiphilic product, depicting the hydrophobic product component, $B$. Micelles nucleate at the center of the simulation cell, move towards the sides  (as indicated by the arrows) and form vesicles, once large enough. At the sides, high competition for product leads to shrinkage of the vesicles. The individual topological changes are depicted below: (b) nucleation of micelle after supersaturation of the solution with product, (c) micelle to vesicle transformation by flip-flopping of amphiphiles to the inside and (d) shrinking of vesicles near the fuel source.}
    \label{fig:vesicles-collage}
\end{figure}

Micelles nucleate at the center of the simulation cell. As they move towards the fuel surface, they gradually grow and a hydrophilic center can form inside the micelle core, facilitated by precursor that has recently reverted. This hydrophilic center then acts as a nucleation site for amphiphiles to flip-flop to the inside and form a vesicle,\cite{hafner_reaction-driven_2023,he_spontaneous_2008} similar to the mechanism of vacuolization of \ac{RDA}-formed droplets in \autoref{sec:implicit_waste}.  The nucleation process to a micelle is depicted in \autoref{fig:vesicles-collage} (b) and the transformation from micelle to vesicle in \autoref{fig:vesicles-collage} (c). Hence, the directed movement inside the fuel-concentration gradient allows to circumvent free-energy barriers and gradually build more complex structures, which would not form otherwise. 

At the fuel source (\ie~simulation cell boundary), the vesicles collide but do not fuse. Since the closely packed vesicles are deprived of their accumulation volume they slowly disintegrate, as seen in \autoref{fig:vesicles-collage} (d).

\section{Conclusion}
\label{sec:conclusion}

In this work, we explored the diffusiophoresis of liquid condensates both, passive and driven by chemical reaction cycles. We showed that the directed movement of droplets within externally maintained concentration gradients differs from diffusiophoresis of colloids, because there exists a finite flux within the droplet. This flux through the droplets can be depleted or enhanced, depending on the interactions of the droplet material, and it dictates the direction of the droplet's movement. By virtue of the incompressibility constraint, the movement is opposed to the flux of fuel or waste through the droplet and is independent of the aggregate size. 

Concentration gradients naturally occur in systems that are driven by chemical reactions due to the local sources of fuel and sinks of waste. For reaction-driven assembly (RDA)-formed droplets, the flux-mediated movement persists. Analytically, we also found additional contributions to the droplet velocity that arise from the asymmetry of the reactions in the droplet's surrounding. These turned out to be corrections, for the cases studied.

Importantly, the fluxes of fuel and waste generated by the chemical reactions will direct the movements of aggregates, through which fuel or waste flow, independently of whether or not they participate in the reaction cycle.

A concentration gradient and the concomitant movement of droplets or aggregates provides a temporal change of environment that modulates the nucleation and growth of aggregates. This facilitates transformations of the aggregate structure, \eg~vacuolization of droplets or the transformation from micelles to vesicles that are otherwise hindered by large free-energy barriers. 

In a broader biological context, this mechanism of directed movement offers a straightforward yet robust and versatile means of transporting organelles, with or without membranes, within finite domains -- akin to (early) cellular systems. For cells we have estimated velocities up to $(10-100)$nm/s. The effect may also guide the design of synthetic reaction-driven assemblies. The direction of transport for these organelles is distinct for each type and sensitively depends on their interactions with fuel and waste constituents.

\section{Methods}
\label{sec:theretical_framework}

Within the continuum model, normalized local concentrations, $\phi_c(\mathbf{r})$ of solvent, reactant~(precursor), product, fuel and waste, $c=S,R,P,F,W$, are taken as order parameters, that determine the free-energy. We employ the Flory-Huggins-de Gennes free-energy functional of \autoref{eq:free-energy-functional}. In the free-energy, the first term is the Flory-Huggins entropy of mixing \cite{flory_thermodynamics_1942,huggins_solutions_1941}, the second one is a square gradient penalty to create a finite interface width. The concentration-dependence of the coefficient of the square-gradient theory is motivated by the Lifshitz entropy, characterizing the narrow interfaces of polymers.
Components interact \textit{via} binary interactions whose strength is determined by the Flory-Huggins parameters $\chi_{cc'}$ in the third term. The system is incompressible, which is enforced by the last term with the Lagrange multiplier, $\pi(\mathbf{r})$.

Given the free-energy functional, we obtain the chemical potentials as the functional derivatives with respect to the concentrations, $\mu_c(\mathbf{r}) = \frac{\delta\mathcal{F}}{\delta\phi_c(\mathbf{r})}$, which yields
\begin{align}
        \frac{\mu_{c}(\mathbf{r})\R^3}{\sqrt{\Nbar}k_{\rm B}T} &= \frac{N_P}{N_{c}} \ln \phi_{c}
        - \frac{\R^2}{6} \frac{\nabla^2 \sqrt{\phi_{c}}}{\sqrt{\phi_{c}}} \nonumber\\
        &+ \sum_{c' \neq c} \chi_{c c'} N_P \phi_{c'} + \pi N_P. \label{eq:chemical_potential}
\end{align}
Gradients in the chemical potentials give rise to fluxes that minimize the free energy,
\begin{align}
    \mathbf{j}_c(\mathbf{r},t) &= - \sum_{c'}\hat{\Lambda}_{cc'}(\mathbf{r},t) \nabla \mu_{c'}(\mathbf{r},t) \nonumber \\
    &+ \sqrt{\phi_c(\mathbf{r},t)}\xi_c(\mathbf{r},t),
    \label{eq:flux}
\end{align}
where the Onsager matrix is diagonal, $\hat{\Lambda}_{cc'}(\mathbf{r},t) = \delta_{cc'}\frac{\lambda_c\R^5}{\sqrt{\Nbar}\kT} \phi_c(\mathbf{r},t)$, since incompressibility is enforced by the free-energy functional. In the following, we define $\Lambda_c:= \hat\Lambda_{cc}$. $\lambda_P R_e^2$ is the diffusion coefficient of the product molecule, $P$. Its self-diffusion time, $\lambda_P^{-1}$, is taken as reference time scale. The factor $\lambda_c$ determines the mobility of component, $c$, which we take the same for all components, $\lambda_c=\lambda_P=:\lambda$. Furthermore, $\xi_c(\mathbf{r},t)$ is Gaussian thermal noise. Its moments are determined by the fluctuation-dissipation relation \cite{van_vlimmeren_calculation_1996,archer_dynamical_2004}
	\begin{eqnarray}
		\langle \xi_c(\mathbf{r}, t)\rangle &=& 0 \\
		\langle \xi_c(\mathbf{r}, t)\xi_{c^\prime}(\mathbf{r}^\prime, t^\prime)\rangle 
		&= &
		\frac{2\lambda}{\sqrt{\Nbar}}\delta_{cc^\prime}\delta(t-t^\prime)\delta(\mathbf{r}-\mathbf{r}^\prime)
		\label{eq:modelB}
	\end{eqnarray}
Without the reactions, concentrations are locally conserved, hence the equilibration follows model-B dynamics \cite{hohenberg_theory_1977}. Meanwhile, insertion of fuel to the system enables reactions that drive the system out of equilibrium. \cite{weber_physics_2019-1,zwicker_intertwined_2022} This yields the time evolution of the concentrations
\begin{align}
	\frac{\partial\phi_c(\mathbf{r},t)}{\partial t} &= \nabla \cdot \left[ \Lambda_c(\mathbf{r},t) \nabla \mu_c(\mathbf{r},t)  + \sqrt{\phi_c(\mathbf{r},t)}\xi_c(\mathbf{r},t)\right]\nonumber\\
 &+ s_{c}(\mathbf{r},t).
    \label{eq:full-time-evolution}	
\end{align}
The reaction rates are $s_{c}=s_{f,c}+s_{b,c}+s_{\mathrm{source},c}+s_{\mathrm{sink},c}$ from forward and backward reaction, source and sink. The first two are given by
\begin{align}
	s_{f,P} = -s_{f,R}  & = -s_{f,F}\frac{N_P}{N_F} = s_{f,W}\frac{N_P}{N_W}\nonumber\\
                        &= r_f N_P\phi_R \phi_F \\
	s_{b,R} = -s_{f,P}  &= r_b N_P\phi_P 
\end{align}
\ie~the forward reaction creates product, diminishes reactant and fuel, and accumulates waste, leading to transient assembly provided that the initial conditions contain fuel. To maintain a nonequilibrium steady state, however, fuel needs to be inserted into the system locally. Here, we consider a cuboidal domain with periodic boundary conditions, where a fuel source and waste sink are located inside slabs at $z=0$ and $z=L_z$. Inside these slabs solvent is replaced by fuel or waste by solvent, respectively,
\begin{align}
	s_{\mathrm{source},F} &= - s_{\mathrm{source},S}\nonumber\\
                            &= r_{\mathrm{source}} \phi_S (\delta (z)+\delta(z-L_z)) \\
	s_{\mathrm{sink},S} &= - s_{\mathrm{sink},W}\nonumber\\
                                &= r_{\mathrm{sink}} \phi_W (\delta (z)+\delta(z-L_z))
\end{align}

Thereby the material conversion of fuel to waste in the bulk is compensated, allowing for the emergence of a statistically stationary nonequilibrium state. 

The Product, $P$, phase separates from solution, forming droplets, whereas the precursor, $R$, is a solute to the solvent. Hence, there is a repulsion of product to solvent and precursor. The backward reaction rates are chosen, such that the droplet radii are only a few $\R$. The forward reaction rate constant is scaled, such that the forward and backward reaction probability for the reactant (precursor) are identical at a small fuel concentration, $\phi_F=0.025$. This justifies the choice of parameters in \autoref{tab:parameter_choice}. 

\begin{table*}
    \centering
    {\small 
    \begin{tabular}{c|c|c|c|c|c|c|c|c|c|c|c}
     $\sqrt{\Nbar}$ & $N_{R/P}$ & $N_{S/F/W}$ & $\chi_{PS}N_P$ & $\chi_{PR}N_P$ & $\chi_{P\,F/W}N_P$ & $r_b$ & $r_f$ & $r_{\mathrm{source}}$ & $r_{\mathrm{sink}}$\\ \hline
     1000 &  10 & 1  & $50$ & $20$ & $0$, $-20$, $-40$ & $(4,8)\cdot 10^{-2}\lambda$ & $40r_b$ & $\frac{r_b\R}{4}$ & $10\lambda\R$
    \end{tabular}
    }
    \caption{Choice of parameters in the Flory-Huggins-de Gennes model. For parameters that are varied, multiple values are given.}
    \label{tab:parameter_choice}
\end{table*}

The size of the simulation cell is chosen $V= 9.6\R \times 9.6 \R \times 25.6\R$, with a spatial resolution of $10$ grid cells per $\R$. We introduce a wall for the droplets, which is modeled \textit{via} an external field. The external field is introduced by adding the term $\sum_c f_{\mathrm{ext},c}(z)\phi_c(\mathbf{r},t)$ to the integrand of \autoref{eq:free-energy-functional}. 
By using a large positive value for precursor and product close to the origin, $z=0$ and $L_z$, we suppress their densities,
\begin{align}
    f_{\mathrm{ext},c} = \left\{ \begin{array}{cl}
        2 & \text{ for }|z| < 0.1\R \\
        1 & \text{ for }0.1 \leq |z| < 0.2\R \\
        0 & \text{ else}
    \end{array}\right.
\end{align}
for $c=R,P$ and $f_{\mathrm{ext},c}=0$ for $c=S,F,W$. This hinders clogging of source and sink by aggregates and collisions of incoming aggregates, with no qualitative effect on the overall dynamics of the system.

All simulations are started from homogeneous initial conditions with mean concentrations of $\bar\phi_R=0.115$, $\phi_S=0.855$, and $\bar\phi_{P,F,W}=0.01$\footnote{We start with very small but non-zero values to avoid numerical instabilities in the continuum model simulations.}. Notice, that the overall concentration of reactant plus product stays constant, $\bar\phi_R+\bar\phi_P=0.125$. If not stated otherwise, we run the simulations up to $t\lambda=1000$ with a time step of $\Delta t\lambda=4\cdot 10^{-5}$, \ie, $2.5\cdot10^7$ time steps. Measurements and averaged are conducted over the last $t\lambda=500$ when the systems have reached a statistically stationary state.

The coupled, partial differential equations are numerically solved using the pseudo-spectral method with semi-implicit time-stepping implemented on a \ac{GPU}, as described in more detail in Ref. \cite{hafner_reaction-driven_2023}. Additional information about the software usage, availability and analysis steps is compiled in \autoref{sec:analysis_procedure} of the \ac{SI}.

\section*{Acknowledgement}
This research was conducted within the Max Planck School Matter to Life supported by the German Federal Ministry of Education and Research (BMBF) in collaboration with the Max Planck Society. The authors gratefully acknowledge the Gauss Centre for Supercomputing e.V. (www.gauss-centre.eu) for providing computing time through the John von Neumann Institute for Computing (NIC) on the GCS Supercomputer JUWELS at Jülich Supercomputing Centre (JSC), as well as the North-German Supercomputing Alliance (HLRN).

\bibliography{rda-concentration-gradients}
\clearpage


\setcounter{section}{0}
\renewcommand{\thesection}{S \Roman{section}}
\setcounter{equation}{0}
\renewcommand{\theequation}{S\arabic{equation}}
\renewcommand{\theHequation}{S\arabic{equation}}
\setcounter{figure}{0}
\renewcommand{\thefigure}{S\arabic{figure}}
\renewcommand{\theHfigure}{S\arabic{figure}}

\begin{widetext}
{\centering\Large \textbf{Supporting Information}}

\section{Theoretical Prediction of Velocities}
\label{sec:velocities_analytical}

In the following, we analytically calculate a prediction for the droplet velocity that takes averaged concentration profiles, as well as locally averaged radii and accumulation volumes from the simulations as input. We start with a \ac{1D} calculation and subsequently extend the result to the \ac{3D} case. 

First, we evaluate the fluxes, satisfying the condition of incompressibility. The flux of component $c$ is given by \autoref{eq:flux} which, ignoring thermal noise, we rewrite to
 \begin{align}
     \mathbf{j}_c(\mathbf{r}) = - \Lambda_c(\mathbf{r}) \nabla \left(\mu_c^0(\mathbf{r}) + \pi(\mathbf{r})N_P\frac{\sqrt{\Nbar}\kT}{R_e^3}\right)
     \label{eq:flux_with_lagrange_multiplier}
 \end{align}
where the bare chemical potentials $\mu_c^0$ are the chemical potentials without the contribution of the Lagrange multiplier \cite{dreyer_simulation_2022,hafner_reaction-driven_2023}. Incompressibility demands the total concentration be constant and hence the total flux vanishes, $\sum_c \mathbf{j}_c = 0$. This allows us to determine the Lagrange multiplier exactly, 
\begin{align}
    \nabla \pi(\mathbf{r})N_P\frac{\sqrt{\Nbar}\kT}{R_e^3}= - \frac{\sum_{c'}\Lambda_{c'}\nabla\mu_{c'}^0}{\sum_{c'}\Lambda_{c'}}.
\end{align}
Plugging this back into \autoref{eq:flux_with_lagrange_multiplier}, we obtain
\begin{align}
    \mathbf{j}_c = \sum_{c'} \frac{\Lambda_c\Lambda_{c'}}{\sum_{\tilde c}\Lambda_{\tilde c}} \nabla \left(\mu_c^0 - \mu_{c'}^0 \right) = \frac{1}{\sum_{\tilde c}\lambda_{\tilde c}\phi_{\tilde c}} \sum_{c'}   \left(\lambda_c'\phi_{c'}\mathbf{j}_c^0 -\lambda_c\phi_{c} \mathbf{j}_{c'}^0 \right),
    \label{eq:flux-lagrange-multiplicator-eliminated}
\end{align}
with the bare fluxes
 \begin{align}
 	\mathbf{j}_c^0 = - \Lambda_c \nabla\mu_c^0 =  - \lambda_c \R^2 \left( \frac{N_P}{N_c} \nabla \phi_c - \frac{\R^2}{6}\phi_c \nabla\left(\frac{\nabla^2\sqrt{\phi_c}}{\sqrt{\phi_c}}\right) + \sum_{c'\neq c} \chi_{cc'}N_P\phi_c \nabla \phi_{c'} \right).
 	\label{eq:bare-flux}
 \end{align}
With this form of coupling the bare fluxes, the incompressibility constraint is fulfilled.

Now we continue with the evaluation of droplet velocities. For a \ac{1D} droplet in concentration gradients, we use the sharp-interface approximation. In this section we consider the dynamics in the vicinity of an individual droplet and omit the index $i$ running over these. We assume that the droplet with radius $R$ positioned at $z_{\rm cm}=0$ at time $t=0$ accumulates product material from its closest surroundings (\ie, \ac{1D} analog of Voronoi cell), $\Omega=[z_{\Omega^-},z_{\Omega^+}]$. Since the droplet moves with velocity, $v$, we exploit that the concentration profiles stays roughly constant in the (instantaneously) co-moving frame, up to small changes upon moving through the concentration gradient, and denote the stationary quantities by a prime, \eg, $\phi_c(z,t) = \phi_c'(z')$ for $z':=z-vt$.  The integral of $\phi_P(z,t)$ over the droplet domain, $\omega=[-R+vt,R+vt]$ yields the total droplet material, which is conserved in the co-moving frame, assuming the droplet size adiabatically adapts to the size that balances loss of product material on the inside and total production in the accumulation volume. Integrating over a domain larger than the droplet domain, then taking the limit yields
\begin{align}
\lim_{\epsilon\rightarrow 0} \int_{-R-\epsilon}^{R+\epsilon} \mathrm{d}z'\, \phi_P'(z') &= \lim_{\epsilon\rightarrow 0} \left[[\phi_P'(z')'z]_{-R-\epsilon}^{R+\epsilon} -\int_{-R-\epsilon}^{R+\epsilon} \mathrm{d}z'\,  \partial_{z'} \phi_P'(z') z' \right]\label{eq:velo_calc_1} \\
 &= \lim_{\epsilon\rightarrow 0} \left[R[\phi_P'(R+\epsilon)+\phi_P'(-R-\epsilon)] + \frac{1}{v}\int_{-R-\epsilon}^{R+\epsilon}\mathrm{d}z'\, \partial_t\phi_P'(z') z' \right]
\end{align}
At this point, we make the approximation (ICA) that product material, which is created inside the accumulation volume instantaneously condensates onto the droplet's interface. Moreover, the droplet shape remains spherical independent of the decay of droplet material on the inside and the condensation on the outside. 

In \ac{1D}, this modifies the time-evolution, \autoref{eq:time-evolution}, for each droplet region, including its accumulation volume, to 
\begin{align}
    \partial_t \phi_P(z,t) = - \partial_z j_P^\mathrm{nr}(z,t) &+ \delta(z+R-vt) \int_{z_{\Omega^-}}^{vt} \mathrm{d}z\, [r_f\phi_R\phi_F(z,t) - r_b\phi_P(z,t)]\zeta(z-z_\mathrm{cm}) \nonumber\\
                &+ \delta(z-R-vt) \int^{z_{\Omega^+}}_{vt} \mathrm{d}z\, [r_f\phi_R\phi_F(z,t) - r_b\phi_P(z,t)]\zeta(z-z_\mathrm{cm})
\end{align}
where $\delta(\cdot)$ denotes Dirac's delta function and $\zeta(z-z_\mathrm{cm})$ is an assignment function, which accounts for the finite lifetime of product molecules. It is estimated for the \ac{3D}-case below. Since the driven reactions are now replaced by surface sources and sinks, we retain only the 'non-reactive' flux, \ie, 
\begin{align}
    j_P^\mathrm{nr}(\mathbf{r}) \approx \left\{ \begin{array}{ll}
        0 & \text{ for }\mathbf{r}\notin \omega \\
        j_P+j_R\approx-(j_F+j_W) & \text{ for }\mathbf{r} \in \omega
    \end{array}, \right.
\end{align}
using that outside the droplet, $\mathbf{r}\notin\omega$, the product concentration almost vanishes $\phi_P\approx 0$.
With this, the above calculation becomes
\begin{align}
    \lim_{\epsilon\rightarrow 0} \int_{-R-\epsilon}^{R+\epsilon}  \mathrm{d}z'\, \phi_P'(z') &= \lim\limits_{\epsilon\rightarrow 0}  {\Bigg [} R[\phi_P'(R+\epsilon)+\phi_P'(-R-\epsilon)] - \frac{1}{v}\int_{-R-\epsilon}^{R+\epsilon}\mathrm{d}z'\, z'  \partial_z {j_P^\mathrm{nr}}'(z') \nonumber \\
 &+ \frac{1}{v}\int_{-R-\epsilon}^{R+\epsilon}\mathrm{d}z'\, z' \delta(z'+R+\epsilon) \int_{z_{\Omega^-}'}^{0} \mathrm{d}z'\,\zeta(z-z_\mathrm{cm}) [r_f\phi_R'\phi_F'(z') - r_b\phi_P'(z')]  \nonumber \\
 &+ \frac{1}{v}\int_{-R-\epsilon}^{R+\epsilon}\mathrm{d}z'\, z' \delta(z'-R-\epsilon) \int^{z_{\Omega^+}'}_{0} \mathrm{d}z'\,\zeta(z-z_\mathrm{cm}) [r_f\phi_R'\phi_F'(z') - r_b\phi_P'(z')] {\Bigg ]}\\
 &= \lim\limits_{\epsilon\rightarrow 0} {\Bigg [}\frac{R}{v}[\underbrace{v\phi_P'(R+\epsilon)-{j_P^\mathrm{nr}}'(R+\epsilon)+v\phi_P'(-R-\epsilon)-{j_P^\mathrm{nr}}'(-R-\epsilon)}_{=0 \,\text{(ICA)}}] \nonumber \\ 
 &- \frac{1}{v}\int_{-R-\epsilon}^{R+\epsilon}\mathrm{d}z'\, [j_F'(z')+j_W'(z')] \nonumber \\
 &- \frac{R}{v} \int_{z_{\Omega^-}'}^{0} \mathrm{d}z'\, \zeta(z-z_\mathrm{cm})[r_f\phi_R'\phi_F'(z') - r_b\phi_P'(z')] \nonumber \\
 &+ \frac{R}{v} \int^{z_{\Omega^+}'}_{0} \mathrm{d}z'\, \zeta(z-z_\mathrm{cm})[r_f\phi_R'\phi_F'(z') - r_b\phi_P'(z')]{\Bigg ]}
\end{align}
where the second term was integrated by parts. Taking the limit, then results in an expression that quantifies the velocity in terms of integrals over the product concentration and the fluxes.
\begin{align}
		v \int_{z_\mathrm{cm}-R}^{z_\mathrm{cm}+R} \mathrm{d}z \,\phi_P(z) =& -\int_{z_\mathrm{cm}-R}^{z_\mathrm{cm}+R} \mathrm{d}z \, [j_F(z)+j_W(z)] \nonumber \\
  &+ \frac{1}{2} \int_{z_{\rm cm}}^{z_{\Omega^+}} \mathrm{d}z\, \zeta(z-z_\mathrm{cm})[r_f\phi_R\phi_F - r_b \phi_P]\nonumber\\
  &-\frac{1}{2}\int_{z_{\Omega^-}}^{z_{\rm cm}} \mathrm{d}z\,\zeta(z-z_\mathrm{cm})[r_f\phi_R\phi_F - r_b \phi_P]. \label{eq:velocity_symmetric_final1d}
\end{align}
There are two contributions: The first arises from the fluxes inside the droplet and the second stems from the reaction/condensation in the droplet's accumulation volume. The second contribution is illustrated in \autoref{fig:reaction_contributions_to_velocity}. 

\begin{figure}
	\centering
	\includegraphics[width=\textwidth]{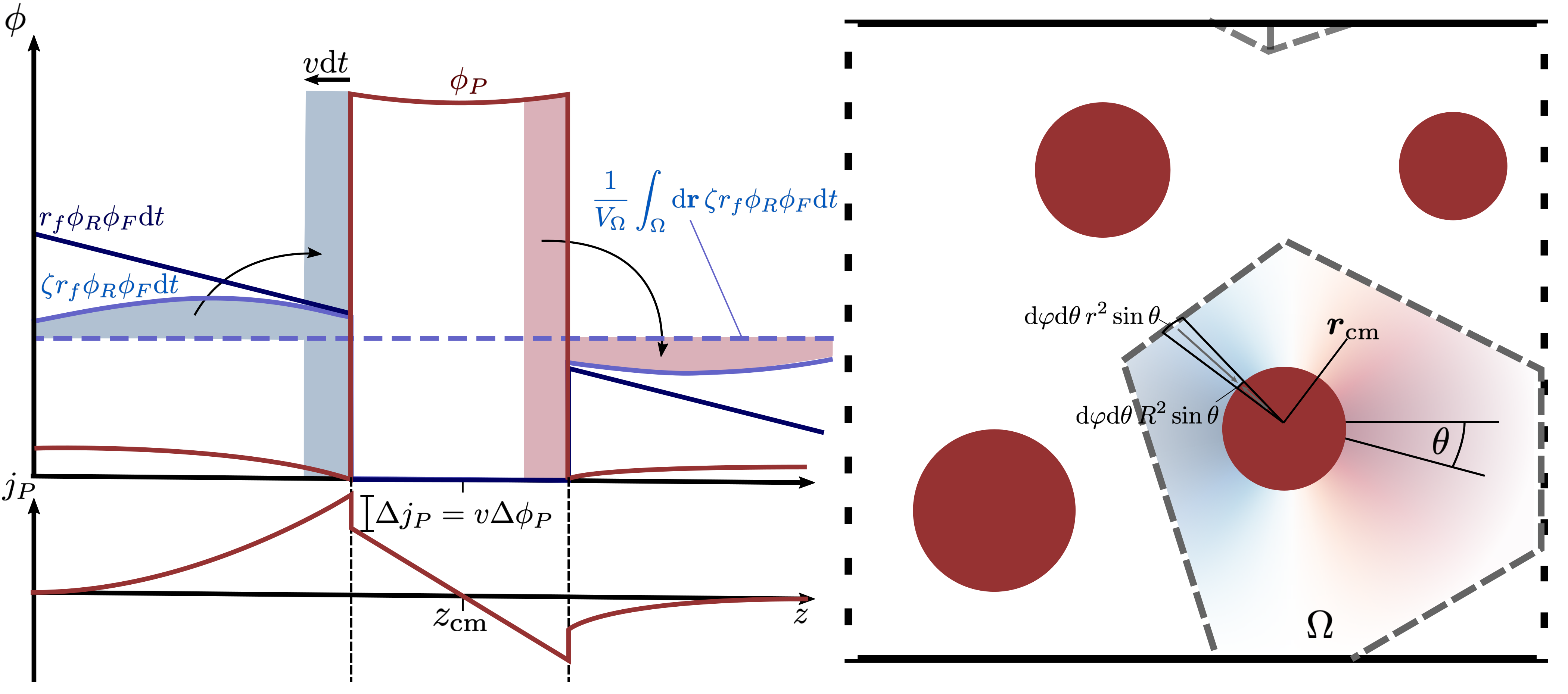}
	\caption{Sketch of the contribution of production effects in fuel and precursor concentration gradients. Left: Visualization of the second contribution of \autoref{eq:velocity_symmetric_final1d}. What is produced as excess to the average in $\Omega$ (for infinitesimal time step $\mathrm{d}t$) leads to movement of the interface, $v\mathrm{d}t$ on the left to the left. Correspondingly, the lack of production compared to the average leads to a retraction of the interface on the right (top). In the corresponding (reactive) product flux profiles there is a jump at the interfaces corresponding to the movement, since there is an excess production on the left (bottom). The reactive flux is set to $0$ in the instantaneous condensation approximation and replaced by surface source and sink shown at the top.  Right: In 3D, droplets (red) are assigned an accumulation volume, $\Omega$, shown by dashed lines which corresponds to the Voronoi domain. The contribution of the local production, $\zeta(|\mathbf{r}-\mathbf{r}_\mathrm{cm}|)\cos(\theta)$, to the velocity in $z$-direction, $v_z=v$ is indicated by the heat map (blue negative, red positive).}
	\label{fig:reaction_contributions_to_velocity}
\end{figure}

Let us generalize the above calculation to \ac{3D}. We assume a spherical droplet of radius $R$ moving at velocity $\mathbf{v} = v \hat{\mathbf{e}}_z$, where $\hat{\mathbf{e}}_z$ is the unit vector, pointing in $z$-direction. The droplet domain is described by $\omega = \{ \mathbf{r} | \; || \mathbf{r} - \mathbf{v} t || < R \}$, and its surface denoted by $\partial \omega$. Again, we approximate all product that is produced within the accumulation volume to condense instantaneously onto the droplet's surface. This modifies the evolution equation to
\begin{align}
    \partial_t \phi_P'(\mathbf{r}') = - \nabla \cdot {\mathbf{j}^{\mathrm{nr}}_P}'(\mathbf{r}',t) &+ \delta(r'-R) \int_{0}^{r_\Omega'(\theta',\varphi')} \mathrm{d}r'\, \frac{r^{'2}}{R^2} \zeta(r') [r_f\phi_R'\phi_F' - r_b\phi_P'](r', \theta', \varphi',t) 
    \label{eq:time-evolution-instant-condensation-3d}
\end{align}
where the co-moving position is $\mathbf{r}'=\mathbf{r}-\mathbf{v}t$, and $r', \theta', \varphi'$ denote the spherical coordinates with respect to the droplet's center. We assume the droplet profile be stationary in the co-moving frame. Here and in the following, all functions of the argument  $\mathbf{r}'$ are denoted by a prime. The fraction that appears in the integral is the result of the fraction of area elements from accumulation in the domain, $r^{'2}\sin\theta' \mathrm{d}\theta'\mathrm{d}\varphi'$, and deposition at the sphere's surface, $R^{2}\sin\theta' \mathrm{d}\theta'\mathrm{d}\varphi'$. Hence, first, we assume that molecules reach the droplet at the angle, at which they have been produced in the accumulation volume. More accurately, the molecules will perform a random motion, before reaching the droplets surface. We will phenomenologically reintroduce this effect below and account for it in the assignment function $\zeta(|\mathbf{r}-\mathbf{r}_\mathrm{cm}|)$. Hence in \ac{3D}, we have

\begin{align}
	\int_{\omega'} \mathrm{d}\mathbf{r}'\, \phi_P'(\mathbf{r}') &= \underbrace{\int_{\partial\omega'} \mathrm{d}\Gamma' \phi_P'(\mathbf{r}') z' (\hat{\mathbf{e}}_{z'} \cdot \hat{\mathbf{e}}_{r'})}_{=:b.t.} - \int_{\omega'} \mathrm{d}\mathbf{r}' z' \underbrace{\hat{\mathbf{e}}_{z'} \cdot \nabla' \phi_P'(\mathbf{r}')}_{=\partial_{z'} \phi_P'(\mathbf{r}-\mathbf{v}t)} \\
	&= b.t. + \frac{1}{v} \int_{\omega'} \mathrm{d}\mathbf{r}' z'  \partial_t\phi_P'(\mathbf{r}') \\
	&\stackrel{\eqref{eq:time-evolution-instant-condensation-3d}}{=} b.t. + \frac{1}{v} \int_{\omega'} \mathrm{d}\mathbf{r}' z' \Bigg[-\nabla'\cdot {\mathbf{j}_P^\mathrm{cm}}'( \mathbf{r}') \nonumber\\
 &+ \delta(r'-R) \int_{0}^{r_\Omega'(\theta',\varphi')} \mathrm{d} r'\, \frac{r^{'2}}{R^2} \zeta(r') [r_f\phi_R'\phi_F' - r_b\phi_P'] \Bigg] \\
	&= b.t. - \frac{1}{v} \int_{\omega'} \mathrm{d}\mathbf{r}' z' \nabla'\cdot {\mathbf{j}_P^\mathrm{nr}}'( \mathbf{r}') +\frac{1}{v} \int_0^{2\pi} \mathrm{d}\varphi' \int_0^\pi \mathrm{d}\theta'\int_0^R \mathrm{d}r' r^{'3} \sin \theta' \cos\theta' \nonumber \\
    &\quad\quad\quad\quad \times\delta(r'-R) \int_{0}^{r_\Omega'(\theta',\varphi')} \mathrm{d} r'\, \frac{r^{'2}}{R^2} \zeta(r') [r_f\phi_R'\phi_F' - r_b\phi_P'] \\
	&= \frac{1}{v} \int_{\partial\omega'} \mathrm{d}\Gamma'\, z' [\mathbf{v}\phi_P'(\mathbf{r}')  - {\mathbf{j}_P^\mathrm{nr}}'(\mathbf{r}')]\cdot \hat{\mathbf{e}}_{r'} + \frac{1}{v} \int_{\omega'} \mathrm{d}\mathbf{r}' \hat{\mathbf{e}}_{z'} \cdot \mathbf{j}'_P( \mathbf{r}') \nonumber \\
	&+\frac{R}{v} \int_0^{2\pi} \mathrm{d}\varphi' \int_0^\pi \mathrm{d}\theta'\int_{0}^{r_\Omega'(\theta',\varphi')}\mathrm{d} r'\, r^{'2} \sin \theta'  [r_f\phi_R'\phi_F' - r_b\phi_P'] \zeta(r') \cos\theta' \\	
	&=  - \frac{1}{v} \int_{\omega'} \mathrm{d}\mathbf{r}' \hat{\mathbf{e}}_{z'} \cdot [\mathbf{j}'_F( \mathbf{r}')+\mathbf{j}'_W( \mathbf{r}')] +\frac{R}{v} \int_{\Omega'}\mathrm{d}\mathbf{r}'\, [r_f\phi_R'\phi_F' - r_b\phi_P']\zeta(r')  \cos\theta' .
\end{align}
where the boundary term vanishes, having formally taken a spherical integration domain with radius  $R\rightarrow R+\epsilon$ with a small $\epsilon>0$ and the subsequent limit $\epsilon \rightarrow 0$, as was explicitly shown in the \ac{1D} case. We obtain the final estimate for the velocity, 
\begin{align}
		v \int_{\omega} \mathrm{d}\mathbf{r}\, \phi_P(\mathbf{r})  &= -\int_{\omega} \mathrm{d}\mathbf{r}\,  [j_F(\mathbf{r})+j_W(\mathbf{r})] + R \int_{\Omega} \mathrm{d}\mathbf{r}\, [r_f\phi_R\phi_F-r_b\phi_P] \zeta(|\mathbf{r}-\mathbf{r}_\mathrm{cm}|)\cos\theta\,, \label{eq:velocity_symmetric_final2}	
\end{align}
where $j_c:=\mathbf{j}_c\cdot\hat{\mathbf{e}}_z$ and again, $\theta$ is the angle between $\mathbf{r}-\mathbf{r}_\mathrm{cm}$ and the $z$-axis, as illustrated in \autoref{fig:reaction_contributions_to_velocity}. The figure also presents a visualization of the second contribution to the velocity. 

Additionally, we account for the random, diffusive motion of molecules before reaching the droplet via the assignment $\zeta(|\mathbf{r}-\mathbf{r}_\mathrm{cm}|)$. Without it, molecules would be assumed to reach the droplet at the angle, where they react to become product. In reality the random motion will blur the position of entry into the droplet. To quantify this factor, we performed simple simulations of a Wiener process, starting at position $\mathbf{r}_0=r \hat{\mathbf{e}}_z$, \ie, $\theta_0=0$ and measured the entry angle $\theta_{\mathrm{arrival}}$ at which the molecules would reach a sphere of radius $R$, located at origin. This allows to calculate the average angle contribution $\langle \cos\theta_{\mathrm{arrival}}\rangle$ and their probability to reach the droplet, $\mathcal{P}_\mathrm{arrival}$. The assignment function is then given by $\zeta(r)=\mathcal{P}_\mathrm{arrival}(r)\langle \cos\theta_{\mathrm{arrival}}\rangle(r)$. The result is shown in \autoref{fig:radial_contribution}.
\begin{figure}
    \centering
    \includegraphics[width=0.8\textwidth]{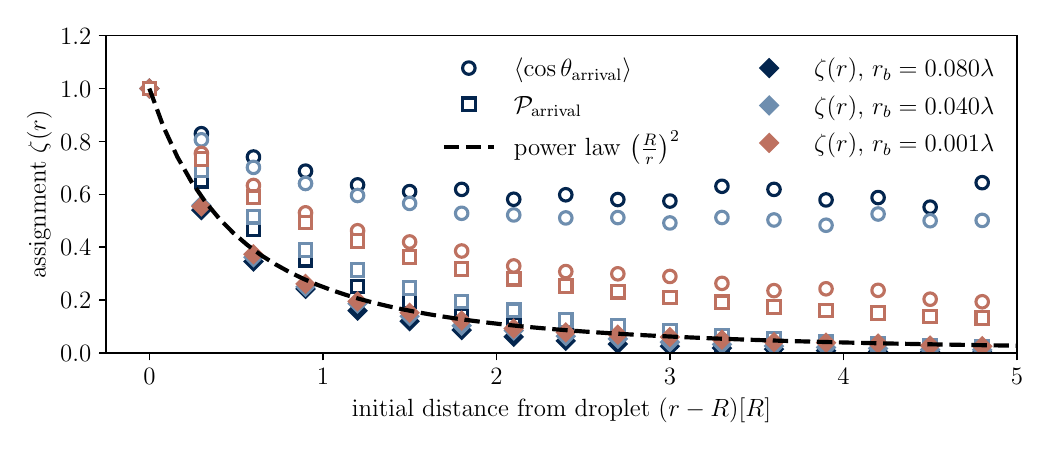}
    \caption{Assignment function, $\zeta(r)$ plotted against distance from the droplet interface, $r-R$, for different reaction rates, which changes the lifetime of product molecules. Additionally, we show the two contributions, arrival probability, $\mathcal{P}_\mathrm{arrival}$ (squares), and directional contribution, $\langle\cos\theta_{\mathrm{arrival}}\rangle$ (spheres). For the different life times the assignment function collapses onto the power law $\zeta(r)=\frac{R^2}{r^2}$ (dashed).}
    \label{fig:radial_contribution}
\end{figure}
Starting within close vicinity of the droplet, molecules have a high probability to reach the surface, $\mathcal{P}_{\mathrm{arrival}}\approx 1$, and do so at an entry angle close to the starting one, $\theta_{\mathrm{arrival}}\approx \theta_0$, \ie, $\cos \theta_{\mathrm{arrival}}\approx 1$. For starting positions further away from the droplet, the arrival probability is lower and the spread of the entry angle is larger, which also diminishes the average $\langle\cos\theta_{\mathrm{arrival}}\rangle$. 
Comparing different reaction rates (corresponding to life times), while the arrival probability is lower for smaller life times, the entry angle, $\theta_{\mathrm{arrival}}$, is spread less such that the assignment functions for all reaction rates (across a wide range of values) collapse onto a universal curve which is nicely described by an effective power law, $\sim r^{-2}$. This motivates the assignment function
\begin{align}
    \zeta(r) = \left\{\begin{array}{cl} 
        1 & \text{for } r<R \\
        \left(\frac{R}{r}\right)^2 & \text{else.}
    \end{array}\right.
\end{align}
Notice that while the assignment function does not explicitly depend on the reaction rate anymore, it still does implicitly because droplet sizes, $R$, are smaller at higher reaction rates.

\section{1D Passive Droplets in External Concentration Gradients}
\label{sec:1DSI}

\begin{figure}
    \centering
    \includegraphics[width=0.8\textwidth]{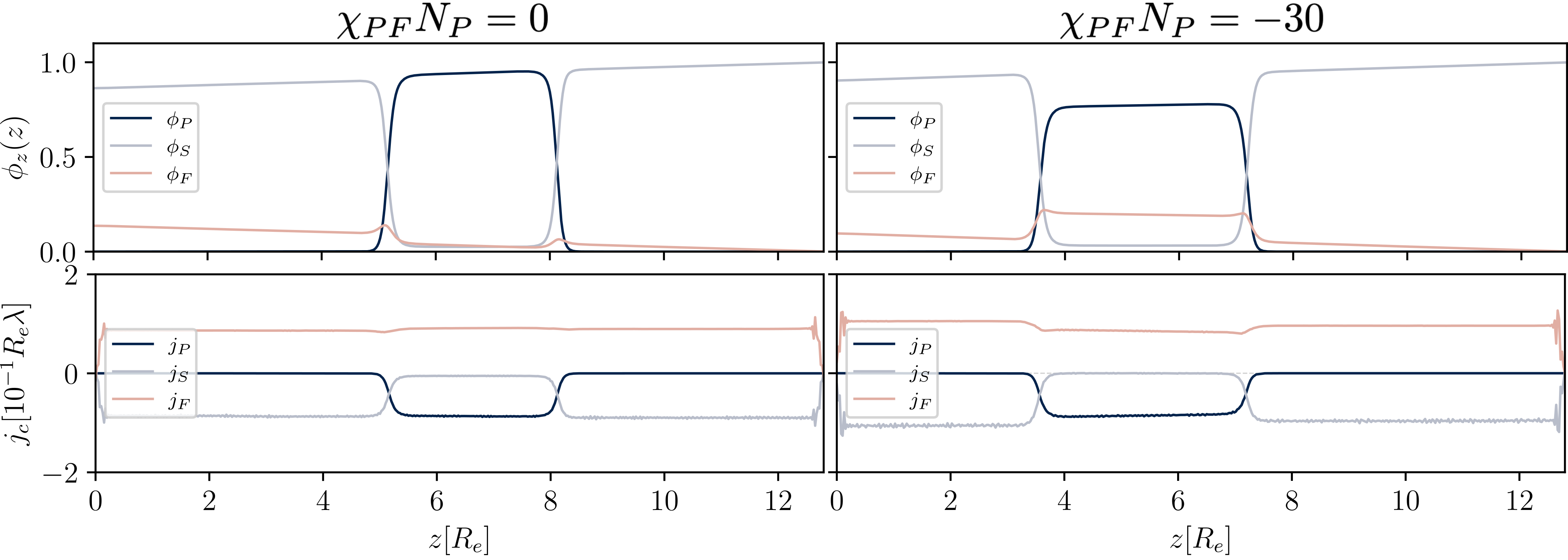}
    \caption{Simulation result of the \ac{1D} passive droplets in external concentration gradients, for two parameter combinations and without thermal noise. In both cases there is a strong repulsion $\chi_{PS}N_P=30$, with neutral or attractive interactions between $F$-solvent and droplet material, as indicated at the top. We show a snapshot at $t\lambda=60$ after placing the droplet at the center of the simulation cell, $z_{\mathrm{cm},t=0} = 12.4\R = \frac{Lz}{2}-0.4\R$. Given are concentration profiles (top) and fluxes (bottom).}
    \label{fig:1dpassive_diffusiophoresis}
\end{figure}
In this section, we compare the \ac{3D} passive droplets in external concentration gradients of \autoref{sec:passive_diffusiophoresis} of the main text to their \ac{1D} counterpart. As we have seen, in \ac{3D} the flux can flow around the droplet in the case where the droplet is unfavorable for the $F$-solvent, or preferentially through the droplet, in the case where the droplet region is attractive for it. This is at the expense of the regions surrounding the droplet laterally, where the $F$-solvent flux is in- or decreased, respectively. In \ac{1D} such a geometric modification of the flux fields is not possible. In such a scenario the flux has to be constant in the bulk regions and jumps at the interface, if it moves with velocity $v$ and there is a jump in concentration, for all components. Indeed this is what we observe in the \ac{1D} simulations. This is shown in \autoref{fig:1dpassive_diffusiophoresis}. Again, the $F$-solvent is depleted inside the droplet for neutral interactions, because of the asymmetry in molecular volume, while it enriched inside the droplet for favorable interactions. This is the same is in the \ac{3D} case. However, the $F$-flux behaves opposite to the \ac{3D} case: It is higher on the inside than on the outside for the depleted concentration and it is reversed for the enriched concentration. This relation is simply dictated by the movement of the droplet, which demands $\Delta j_F = v\Delta \phi_F$. This explains the higher $F$-flux inside the droplet for the case $\chi_{PF}N_P=0$ than on the outside (at the left interface: $\Delta \phi_F \approx -0.05$, $v\approx-0.1\R\lambda$, $\Delta j_F\approx 0.005\R\lambda$), and correspondingly the lower one inside for $\chi_{PF}N_P=-30$ ($\Delta \phi_F \approx 0.14$, $v\approx-0.12\R\lambda$, $\Delta j_F\approx -0.018\R\lambda$). Compared to the \ac{3D} case, here, all material of $F$ and $S$ necessarily must flow through the droplet. Thus, the droplet is a bottleneck in this scenario and influences the (otherwise constant)\footnote{Note that the movement of the droplet, with other $F$-concentrations than in the surroundings implies a change of  mean concentration in the system that changes the concentration profiles surrounding it and hence the flux is not strictly constant but skewed. This is slightly visible in \autoref{fig:1dpassive_diffusiophoresis} with opposite skewedness for the two cases.} flux in the whole simulation cell. That way the droplet is still a little bit faster in the attractive scenario than in the neutral one.

\section{Velocity of Passive Droplets in External Concentration Gradients}
\label{sec:velocity_passive_droplet}
\begin{figure}
    \centering
    \includegraphics[width=0.5\textwidth]{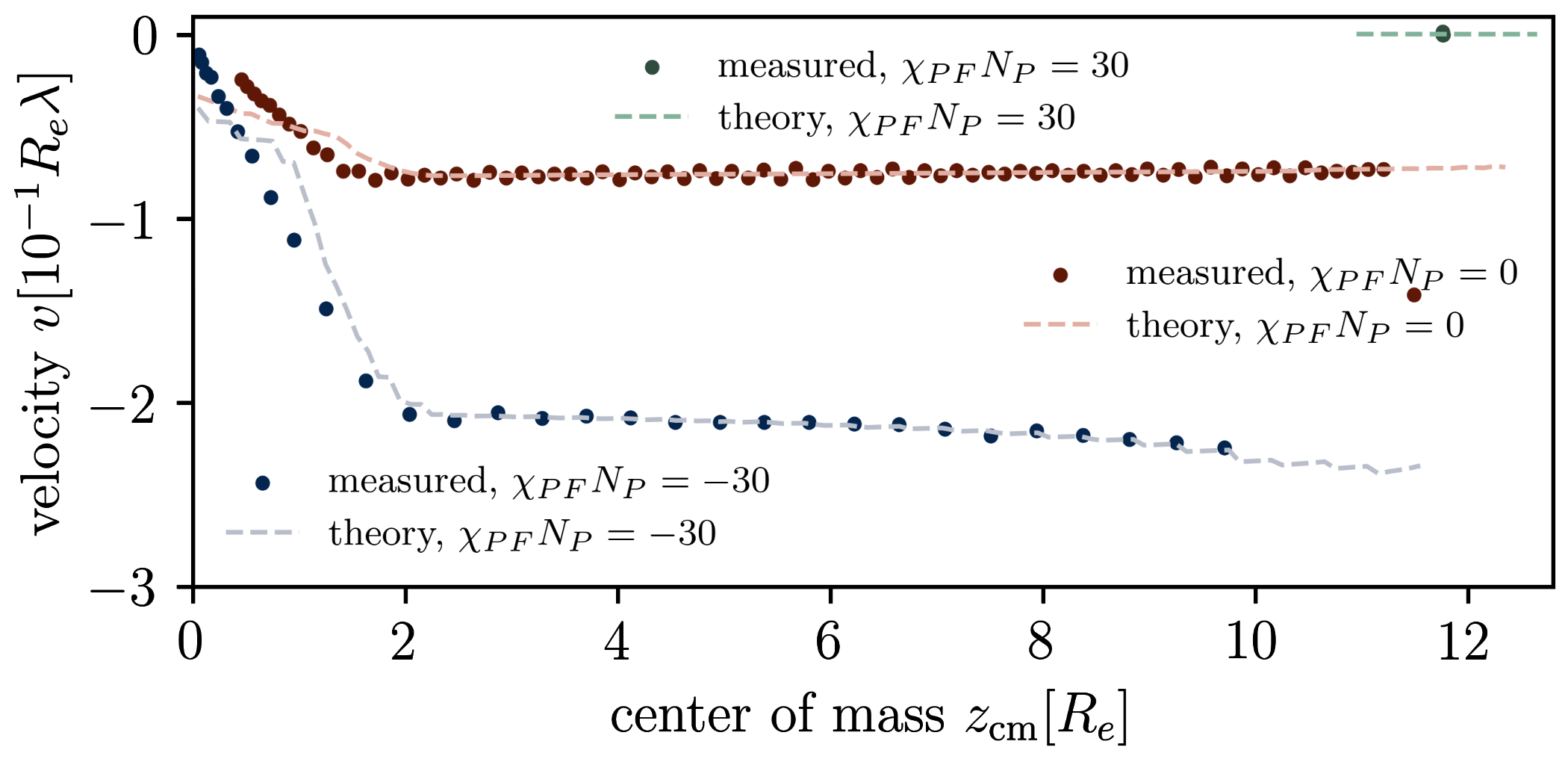}
    \caption{Droplet velocity in $z$-direction, $v$, plotted against center of mass, $z_{\mathrm{cm}}$, both as measured from simulation and with the theoretical prediction using the measured fluxes and concentrations inside the droplet.}
    \label{fig:passive-droplet-velocity}
\end{figure}
For the simulations of \autoref{sec:passive_diffusiophoresis} we can measure the droplet velocity and compare it the theoretical prediction above, which relates it to the fluxes and concentrations.

To do so, we perform the \ac{HKCA} \cite{hoshen_percolation_1976,hoshen_percolation_1997}, where a droplet is defined by the threshold $\phi_P>0.5$. This allows to measure the droplets center of mass, $\mathbf{r}_{\mathrm{cm}}$, and thereby its velocity using two consecutive time steps. \autoref{fig:passive-droplet-velocity} depicts the measured velocities as a function of the center of mass. The velocities are approximately constant inside the bulk and approach zero once the droplet overlaps with the $F$-source, \ie, once its center of mass reaches $z=R$, with $R$ being the droplet's radius. We compare the measured velocities to the theoretical prediction, which is worked out for the general case of \ac{RDA}-formed droplets in fuel and waste concentration gradients in \autoref{sec:velocities_analytical}, resulting in \autoref{eq:velocity_symmetric_final2}. For the chemically passive droplets, $r_f=r_b=0$, and assuming the concentration profiles vary slowly across the droplet volume, this simply reduces to $v=\frac{j_P(\mathbf{r}_{\mathrm{cm}})}{\phi_P(\mathbf{r}_{\mathrm{cm}})}$. Making use of the cluster analysis, we average the fluxes and concentrations within the droplets at every time step and use these averages to predict $v$. The result is plotted in \autoref{fig:passive-droplet-velocity} as dashed lines and matches the measurements. 

\section{Nucleation of RDA-formed Droplets}
\label{sec:nucleation}
\begin{figure}
    \centering
    \includegraphics[width=\textwidth]{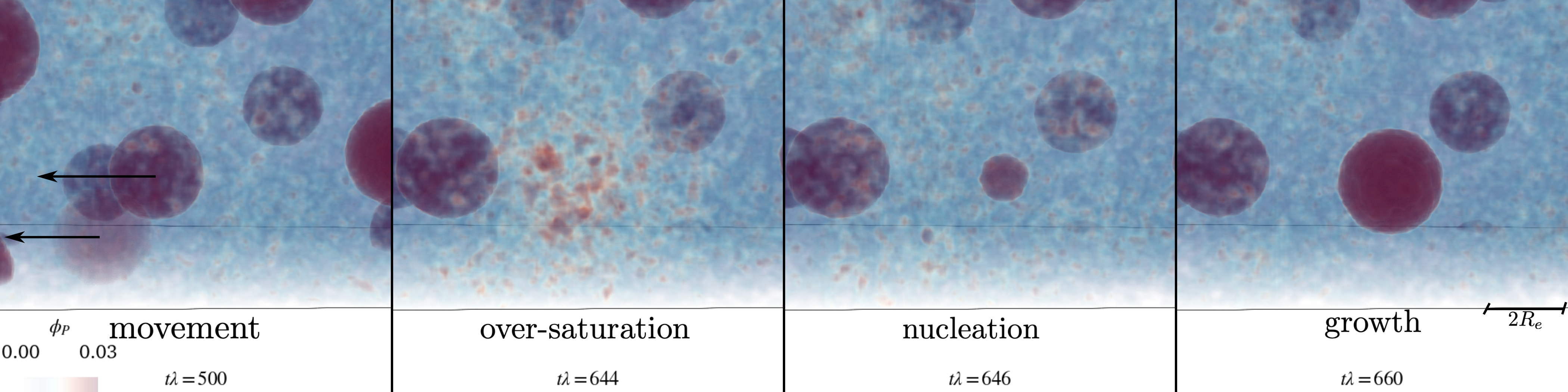}
    \caption{Nucleation dynamics of \ac{RDA}-formed droplets for the setup with implicit waste of \autoref{sec:implicit_waste} in the main text. The different dynamics are indicated.}
    \label{fig:nucleation}
\end{figure}
Here, we briefly demonstrate how \ac{RDA}-formed droplets nucleate. We show this on the example system of \autoref{sec:implicit_waste} of the main text. The droplets move towards the sides of the simulation cell, towards higher fuel concentrations. In this setup, the collective movement creates large voids between droplets in which the forward reaction commences. This creates more product in the solution than can diffuse into the surrounding droplets or can revert into the precursor state. Thus, the product concentration in these void begins to rise until it comes close to the spinodal concentration. When this happens, a droplet can stochastically nucleate and such a nucleation event becomes the probable the higher the super-saturation $\phi_P(\mathbf{r})-\phi_{P,\mathrm{binodal}}$. Eventually, a new droplet nucleates, which quickly adsorbs the super-saturated product from its surrounding and grows to the stationary size. Such a nucleation event is shown in \autoref{fig:nucleation}.

Note that the nucleation happens close to the (mean-field) spinodal, \cite{cahn_free_1959}
 \ie, when the free-energy barrier for the nucleation becomes very small (\ie~on the order of $\kT$). The binodal was measured at $\phi_{P,\mathrm{binodal}}=0.002$ (with noise) and the mean-field spinodal at $\phi_{P,\mathrm{spinodal}}=0.01$. This is where we typically observe the nucleation process, \ie~at high supersaturation. Hence, formation of droplets occurs in our simulations at the transition between (classical) nucleation and spinodal decomposition.

\section{Choice of Boundary Conditions}
\label{sec:choice_of_boundary_conditions}

\begin{figure}
    \centering
    \includegraphics[width=\textwidth]{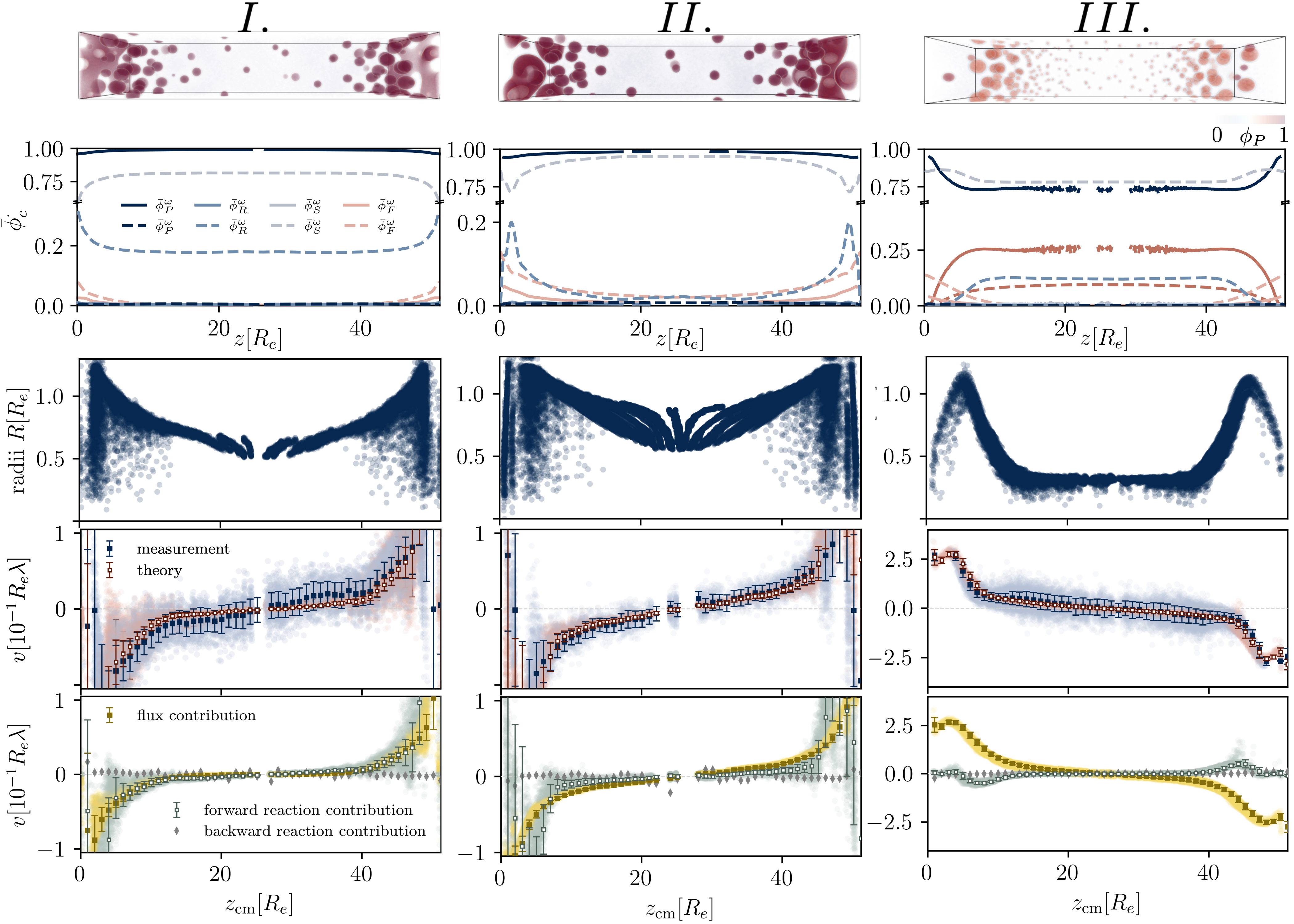}
    \caption{Simulation results for various systems, numbered as described in the text. For each column, the top panel shows an example snapshot of the product concentration field. In the second row, we show the temporally and laterally averaged concentration profiles inside (solid) and outside (dashed) of the droplets. The third row displays the measured droplet radii, plotted against their center of mass. The fourth row shows measured and theoretically predicted velocities plotted against center of mass. Individual measurements are shown as dots and the local averaged is marked with errorbars, indicating the standard deviation of the local distribution. Finally, the fifth panel does the same for the three contributions of the theoretical prediction.}
    \label{fig:system-archs}
\end{figure}
In this section, we discuss different simulation-cell architectures and the resulting dynamics. We are considering three scenarios that differ from the simulations in the main text and compare these. The three scenarios are 
\begin{enumerate}[label=\Roman*.]
    \item Implicit treatment of waste with fully periodic boundary conditions, \ie, no external field at the boundaries of the simulation cell.  Additionally, we take a large $z$-dimension, $L_z=51.2\R$, and double the initial precursor concentration, $\bar\phi_R +\bar\phi_P=0.25$. The missing external field allows for droplets to cover the source of fuel, and thus a larger source rate, $r_{\mathrm{source}}=0.1\R\lambda$, is necessary.
    \item Implicit treatment of waste with larger simulation-cell size, $L_z=51.2\R$, but external boundary field. The larger size of the simulation cell justifies a doubled source rate, $r_{\mathrm{source}}=0.02\R\lambda$. All other parameters are kept the same as in the main text.
    \item Explicit fuel and waste with an affinity of product towards waste, $\chi_{PW}=-40$, and a larger simulation-cell size of $L_z=51.2\R$. All other parameters are unchanged compared to the main text.
 \end{enumerate}

The simulation results for the three system are shown in \autoref{fig:system-archs}. For system I, implicit fuel with a large simulation cell, more precursor and no wall, we observe the nucleation in the center of the system and subsequent directed movement towards the closest fuel source. Instead of stopping in front of the source, droplet freely move to the sides of the simulation cell, where they fuse with oncoming droplets. This allows fusion to large structures that regularly reach beyond the periodic boundary conditions. Allowing the movement above the source also allows to block it which periodically leads to depletion of fuel in the system, reducing the product concentration until the source is free, again. This undesirable dynamics justifies the external boundary field that was applied in all other simulations. Additionally, the large simulation cell allows nucleation of droplets far away from the source. Here the droplets do not grow as quickly as in the case of a smaller simulation cell. Droplets stay small while they are far away from the source and only gradually grow upon approaching it. This becomes visible when looking at the droplet radii plotted against their center of mass. From the laterally averaged concentrations inside and outside of the droplets we can see that the precursor concentration is high in all of the solution. Measuring the velocity as a function of the center of mass results in a similar curve as the one from the main text, nicely matched by the theoretical prediction. Looking at the individual contribution to the prediction, the bottom panel, we realize that in this scenario, the reaction contributions are just as large as the flux contributions.  

System II is closer to the system of the main text, with the difference of a doubled simulation-cell size and doubled refilling rate of the fuel, while we keep the polymer concentration low, $\bar\phi_R+\bar\phi_P=0.125$. Due to the larger size of the simulation cell, we observe, again, a region of small droplet density in the center, where droplets that do nucleate stay small and gradually grow upon approaching the source. In comparison to system I, the precursor concentration becomes low in the center, while the fuel concentration does not fall off as fast. Compared to the smaller system size of the main text, the product is more concentrated at the walls, which results in larger structures that also regularly span beyond the periodic boundaries laterally. Again, the velocity prediction matches the measurements. In this scenario the reaction contribution is comparable to the flux contribution, however smaller in the bulk. 

System III features fuel and waste explicitly, with an affinity of product towards waste. This scenario is tailored to show what happens if the simulation cell is larger than the length scale of the waste concentration profile and thus too large for droplets to approach in the center. As visible in the example snapshot and the radii plot of \autoref{fig:system-archs}, droplets nucleate close to the fuel source. From here they quickly move towards the center. At approximately $z\approx 10\R$ and $z\approx L_z-10\R$ multiple droplets are closely packed, in full correspondence to the scenario with small simulation-cell size. When droplets approach this region, their accumulation volume becomes asymmetric, with small volume closer the center and large volume closer to the source. This leads to large forward reaction contributions opposing the flux contribution, and the droplets are slowed down. In this region the droplets adopt the largest sizes, while they shrink as they move forward from here farther away from the source. After moving through the densely packed region, the droplets distance from each other while flux contribution drives them towards the center. The droplets continuously shrink until they reach their smallest possible size, dictated by the interface width ($R\approx 0.3\R$), and at this point dissolve. 
In this scenario, the velocity measurements also display the 'shoulder', the curve obtaining a small gradient in the center. Here it is not only a result of the jamming and the reaction contribution opposing the flux contribution but also the flux contribution itself displays this behavior because the waste concentration becomes saturated with a vanishing gradient at the center.

These three scenarios demonstrate the robustness of the mechanism, which qualitatively does not depend on the boundary conditions of the simulation cell. Quantitatively, however, the dynamics depends on those details offering opportunities to tailor the behavior.

\section{Reaction Contribution to Velocities}
\label{sec:reaction_contributions}
\begin{figure}
    \centering
    \includegraphics[width=0.8\textwidth]{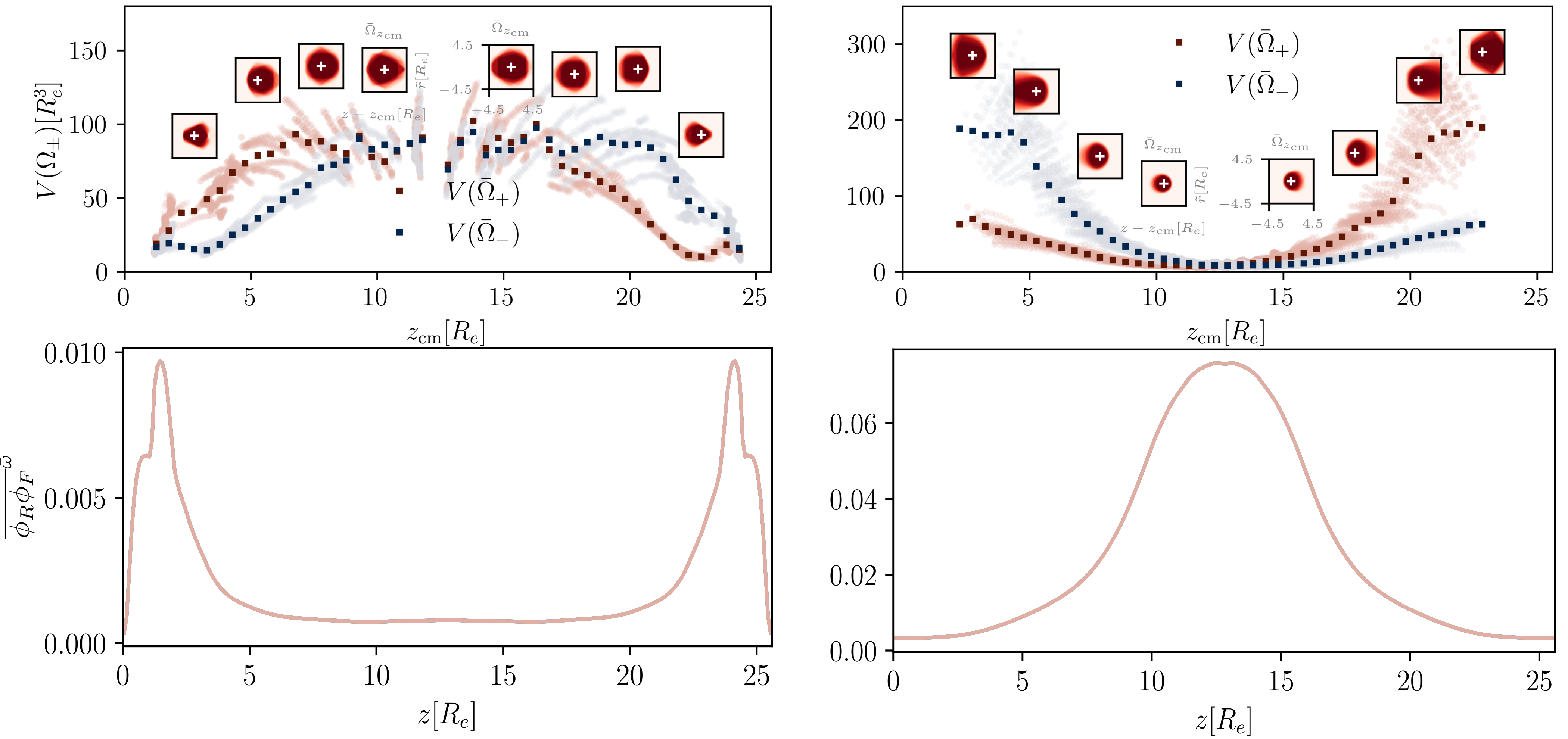}
    \caption{Measurement results for the system of implicit waste (left) and implicit fuel (right). In the top row, we show the mean volume of the accumulation domains in front of, $V(\Omega_-)$, and behind, $V(\Omega_+)$, the center of mass. The insets show the (locally) averaged domains, assuming cylindrical symmetry ($\tilde r = \sqrt{x^2+y^2}$). The center of mass is marked as a cross to see the horizontal asymmetry in these. In the bottom row we show the temporally and laterally averaged product of concentrations $\phi_R\phi_F$, which are proportional to the product of the forward reaction (proportionality factor is, $r_fN_P=3.2N_P\lambda$ on the left and $r_fN_P=0.32N_P\lambda$ on the right).}
    \label{fig:reaction_contr_contr}
\end{figure}
We have shown in the main text that the forward reaction contribution strongly varies, depending on the system's morphology, \ie, the arrangement of droplets and the distribution of precursor and fuel molecules. The movement that can be produced by the forward reaction is the result of an asymmetry in product that flows into the droplet. This can stem either from concentration gradients of fuel and precursor (\ie~their products) or from an asymmetry in the accumulation volume.

For the two examples of implicit waste or fuel of  \autoref{sec:implicit_waste} and \ref{sec:implicit_fuel} of the main text, let us have a look at the statistics of the accumulation volumes and the averaged concentrations. For this we perform the Voronoi tessellation of the domains for each time step and locally average the distinct domains. Assuming cylindrical symmetry of the averaged domains, the result is presented in the insets of \autoref{fig:reaction_contr_contr}. Clearly, the accumulation volumes are asymmetric around the center of mass of the droplets: For implicit fuel and movement towards the sides of the simulation cell, the accumulation volumes are larger farther away from the source. In the opposite case, implicit fuel and movement towards the center, the accumulation volumes are larger at the side that is closer to the waste sink. This motivates analyzing the volume of the accumulation domains in front of the center of mass, $V(\Omega_-)$ for $z<z_{\mathrm{cm}}$, and behind it, $V(\Omega_+)$ for $z>z_{\mathrm{cm}}$. The result is depicted in the top row of \autoref{fig:reaction_contr_contr}.
The bottom row shows the temporally and laterally averaged product of fuel and precursor concentrations outside of the droplets, $\overline{\phi_R\phi_F}^{\bar\omega}$. In the case of implicit waste and movement towards the side, the precursor concentration is also higher at the sides of the simulation cell, as is the fuel concentration, and there is a high gradient of the product concentration. Hence, within a infinitesimal volume around droplet, the production is higher closer to the source than further away from it. It is the opposite for the scenario where droplets move towards the center, although the fuel  concentration (solvent in this scenario) still is highest at the sides. Hence the two contributions to the reaction contribution oppose each other. The result, the green lines of Figures \ref{fig:hp_fg_velocities} and \ref{fig:wg_velocity_collage}, is that the reaction contribution works in the direction of the flux contribution in the first case, while it works in opposite direction of it (away from the center) in the case where droplets move away from the waste sink.

These two scenarios are representative also of the full systems, where both, fuel and waste, are treated explicitly.

\section{Fuel and Waste Fluxes Cancel in the Symmetric Case}
\label{sec:symmetric_interactions}

When interactions and mobilities of fuel and waste are the same, their effects on the droplets cancel exactly in the statistically stationary state, such that the droplets do not move.

For this argument, consider a system where no phase separation occurs. For simplicity we only consider fuel, waste and solvent, treating precursor and product implicitly, and analyze the resulting solvent flux. We take all mobilities to be the same, $\lambda_c=\lambda$ $\forall c$ and all interactions to vanish, $\chi_{cc'}=0\,\forall c,c'$. Moreover, we disregard thermal fluctuations. For the fuel component, the time evolution becomes 

\begin{align}
	\partial_t \phi_F &= \frac{\lambda R_e^5}{\sqrt{\Nbar}\kT} \nabla \cdot \left[ \phi_F\phi_S \nabla (\mu_F-\mu_S) + \phi_F\phi_W \nabla (\mu_F-\mu_W) \right] - r_f(\mathbf{r}) \phi_F \\
	&= \lambda R_e^2 \nabla^2 \phi_F - r_f(\mathbf{r}) \phi_F
	\label{eq:fwg_1d_examplecalcF}
\end{align}
where $r_f(\mathbf{r}) \stackrel{\wedge}{=}r_f \phi_R(\mathbf{r})$ is now a general local forward reaction rate, and in the second line we consider only small concentration gradients, allowing us to disregard the square-gradient penalty in the chemical potential. The time evolution of the waste is analogous, 
\begin{align}
	\partial_t \phi_W = \lambda R_e^2 \nabla^2 \phi_W + r_f(\mathbf{r}) \phi_F.
		\label{eq:fwg_1d_examplecalcW}
\end{align}
In the stationary state, the resulting solvent flux takes the form
\begin{align}
	\mathbf{j}_S = - (\mathbf{j}_F+\mathbf{j}_W) = \lambda R_e^2 \nabla (\phi_F+\phi_W) \stackrel{\eqref{eq:fwg_1d_examplecalcF},\eqref{eq:fwg_1d_examplecalcW}}{=} 0.
\end{align}
Since waste and fuel flux cancel exactly, the main contribution to the droplet velocity vanishes.

Indeed, such a system shows an arrest in movement and approaches a stationary state where large droplets arrange on a lattice structure close to the interface and droplet sizes decrease further away from the source. Here, the classical size-control mechanism of such \ac{RDA} systems becomes visible, dictating the lattice spacing, and size of droplets then being determined by the (local) mean concentration of product, which is smaller further away from the fuel source. The simulation results and analysis are shown in \autoref{fig:fwg_cancelation}. While the flux inside the droplets vanishes entirely, the reaction contributions do not, \ie, the asymmetry in accumulation volumes does not entirely counter the asymmetric production because of concentration gradients.
Here, the deviation between prediction and measurement is again prominent close to the source.

\begin{figure}
    \centering
    \includegraphics[width=0.5\textwidth]{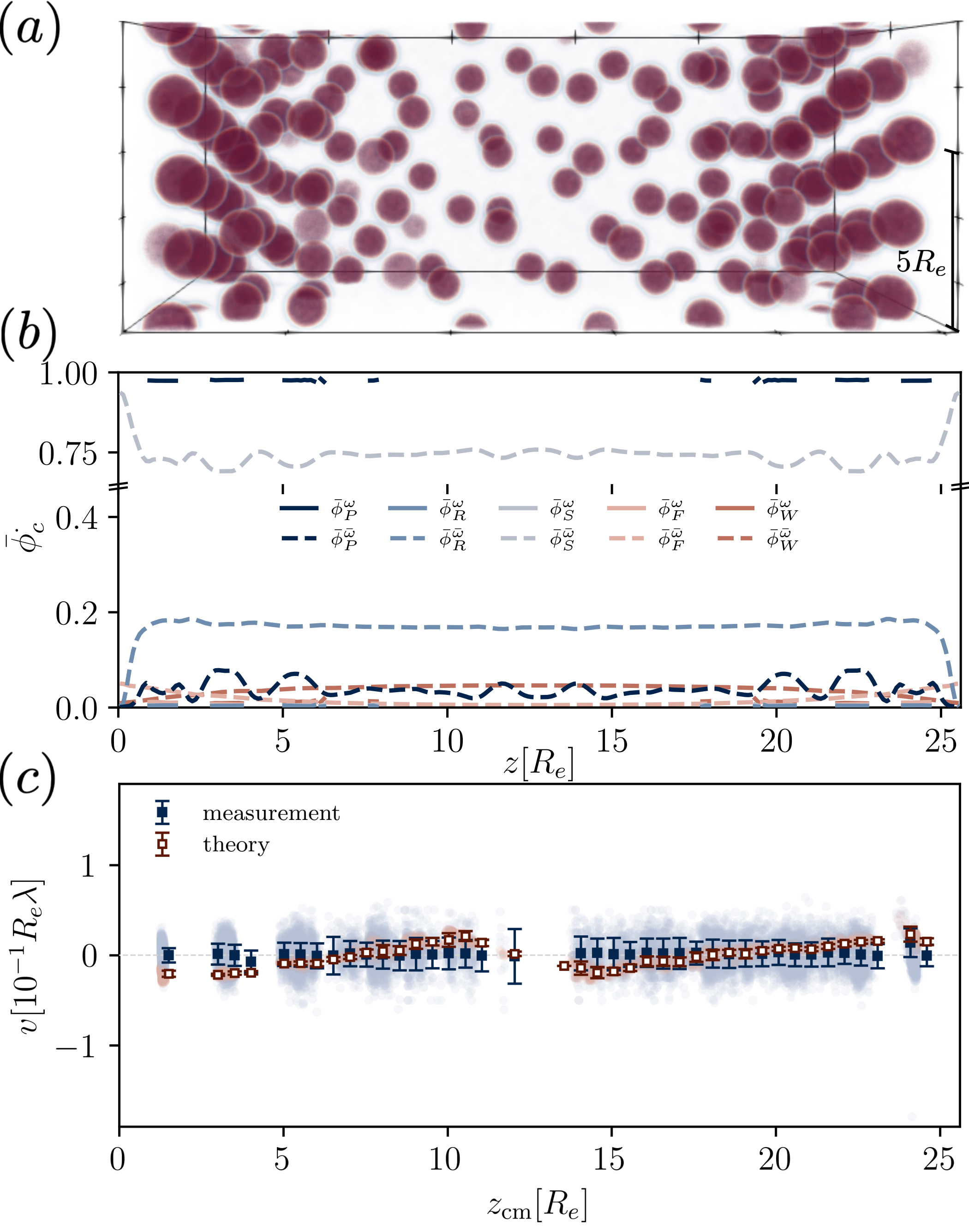}
    \caption{Simulation results for explicit treatment of fuel and waste with symmetric interactions, $\chi_{PF}=\chi_{PW}=0$, showing (a) an example snapshot after time $t\lambda=1900$, after the system has become stationary. (b) The averaged concentration profiles inside and outside the droplets. Due to the lattice structure, in between the droplet layers, there is also no data on the inside concentrations. (c) Measured and theoretical velocities plotted against center of mass.}
    \label{fig:fwg_cancelation}
\end{figure}

\section{Simulation Results for Explicit Fuel and Waste}
\label{sec:simulation-results-explicit}
\begin{figure}
	\centering
	\includegraphics[width=\textwidth]{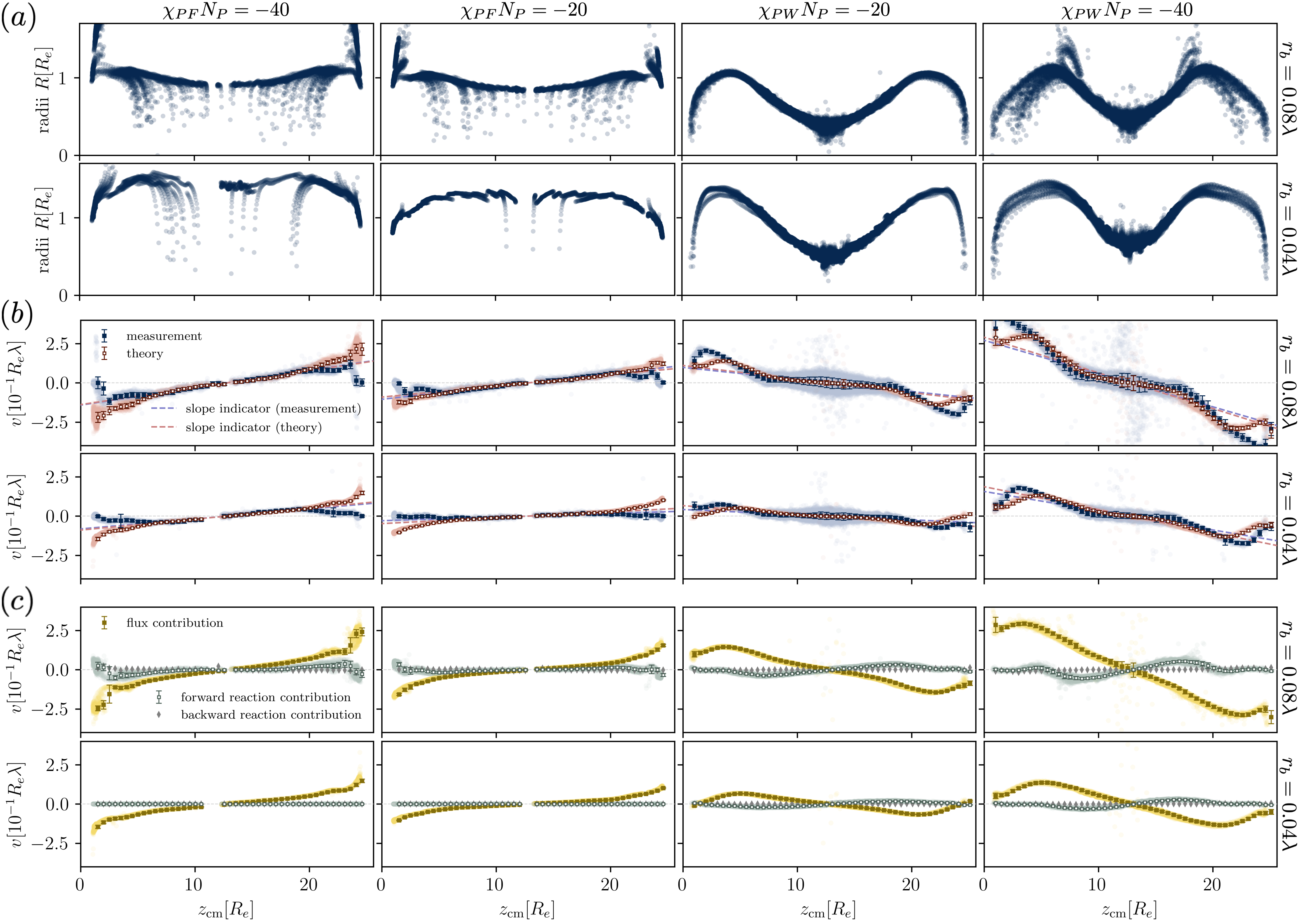}
	\caption{Simulation results for explicit treatment of fuel and waste for multiple system parameters, varying fuel and waste interactions with the product, $\chi_{PF}N_P$ and $\chi_{PW}N_P$, along the columns, and reactions rates, $r_b$, along rows. Among the interactions, one interaction (not indicated in the figure) is set to $0$ (nonpreferential).  Each quantity is plotted against the droplet center of mass, $z_\mathrm{cm}$. (a) Measured droplet radius, $R$. (b)
 Measured and theoretically computed droplet velocity, $v$, using \autoref{eq:velocity_symmetric_final}. A linear fit in the bulk, $5\leq z\leq 21.6\R$, is shown for each curve. (c) Dissection of theoretical velocity into its three contributions. In (b) and (c) the colors and markers are the same as in \autoref{fig:hp_fg_velocities}.}
	\label{fig:fwg_velocity_vs_position}
\end{figure}

For explicit treatment of fuel and waste we vary the interactions of fuel and waste with the product, $\chi_{PF}$ and $\chi_{PW}$, and the reaction rate constants, $r_b$. Performing all measurements, as before, we obtain the simulation results in \autoref{fig:fwg_velocity_vs_position}. Panel (a) shows the measured droplet radii plotted against the droplets' center of mass. Qualitatively one can observe nucleation in the center for the attractive-fuel case. $\chi_{PF}<0$, while the nucleation happens near the source in the opposite case, $\chi_{PW}<0$. For large affinities one can also observe fusion events where the droplets move to, while such events are not observable for intermediate attractions (lower velocities). The droplets become larger for smaller reaction rates, $r_b=0.04\lambda$. The velocity measurements and their theoretical comparisons are plotted in panel (b). Clearly strong attractions lead to higher velocities, as well as a higher reaction rate, which is associated with higher concentration gradients. The linear fit in the bulk, $5\leq z\leq 21.6\R$, is indicated as dashed lines for both measurement and theory. Its slope is used as the indicator for the droplet dynamics in the main text. Panel (c) dissects the theoretical comparison into the three contributions and we obtain the same results, as in the cases of implicit treatment. The flux contribution determines the overall dynamics of the droplets which becomes stronger, the more attractive one of the component becomes. This determines the arrangement of droplets and how these share the product that is created in the surrounding solution. That way, the forward reaction contribution is enslaved to the flux contribution and acts as a small correction to the velocity. The backward reaction contribution is always negligible. 

\section{Particle-based Simulations}
\label{sec:particle-based}

To corroborate the continuum model results, we perform particle-based simulations using a soft, coarse-grained model. We use the Monte-Carlo simulation program \ac{SOMA} \cite{schneider_multi-architecture_2019} which employs the \ac{SCMF} algorithm \cite{daoulas_single_2006}. The model treats molecules as Gaussian chains, with a discretization $N_i$, lumping several monomer repeat units into a single interaction center and employs a top-down approach to assess the interactions and dynamics. The model allows for great parallelization by splitting interactions into strong bonded interaction and weak non-bonded ones, $\mathcal{H} = \mathcal{H}_{\mathrm{b}}+\mathcal{H}_{\mathrm{nb}}$ with 
\begin{align}
	\frac{\mathcal{H}_\mathrm{b}}{\kT} = \sum_m\sum_b \frac{3(N_P-1)}{2\R^2} (\mathbf{r}_{m,b}-\mathbf{r}_{m,b+1})^2,
\end{align}
where $m$ runs over all molecules, $b$ over all bonds within the molecule, and $\mathbf{r}_{m,b}$, $\mathbf{r}_{m,b+1}$ refer to the positions of the bonded particles. The weak non-bonded interactions are expressed in terms of the normalized densities $\phi_c(\mathbf{r})$, \ie, 
\begin{align}
	\frac{\mathcal{H}_{\mathrm{nb}}}{\sqrt{\Nbar}k_{\rm B}T} =
     \int \frac{\mathrm{d}\mathbf{r}}{\R^3} 
    &\left(
    \frac{\kappa_0 N_P}{2} \left[\sum_{c}
    \phi_c(\mathbf{r}) - 1\right]^2 + \frac{1}{2} \sum_{c \neq c^\prime}
    \chi_{cc^\prime}N_P \phi_c(\mathbf{r}) \phi_{c^\prime}(\mathbf{r}) 
    \right),
    \label{eq:non-bonded-energy}
\end{align}
where $\kappa_0=6$ is the inverse isothermal compressibility which is large for the nearly incompressible system. A more detailed description of the particle-based model including the implementation of reactions is given in Refs. \cite{schneider_multi-architecture_2019,hafner_reaction-driven_2023}. 

Conveniently, the two models operate on the same set of parameters which simplifies comparison between the two models. In the particle-based simulations, we take the same parameters, with the exception that solvent, fuel and waste molecules are dimers, $N_{S/F/W}=2$, and the polymer length scales correspondingly, $N_{R/P}=20$, just as was done in \cite{dreyer_simulation_2022,hafner_reaction-driven_2023}. We take simulation cells of  $V=10\R\times 10\R\times 25\R$ at a resolution of $8$ grid cells per $\R$. To comply with this, we also take a higher discretization of the precursor and product molecules, $N_{R/P}=20$. We take a reaction rate constant of $r_b=0.08\tau_0^{-1}$ and vary the fuel and waste interactions with the product molecules. Since product and short fuel and waste molecules have different lengths, the invariant interaction parameter is $\chi_{P,F/W}$ rather than $\chi_{P,F/W}N_P$ as was the case if both species had the same discretization. Therefore, we vary the interactions in the range $\chi_{P,F/W}=0,-2,-4$ to match the interaction of the continuum model with half the discretization values. The results are presented in \autoref{fig:soma_results}, where we depict the temporally and laterally averaged concentrations in- and outside the droplets in (a) and the measured velocities in (b). Visibly, we obtain the same result as before. The affinity towards one component determines the direction of the movement which can be amplified by high affinities.

\begin{figure}
    \centering
    \includegraphics[width=\textwidth]{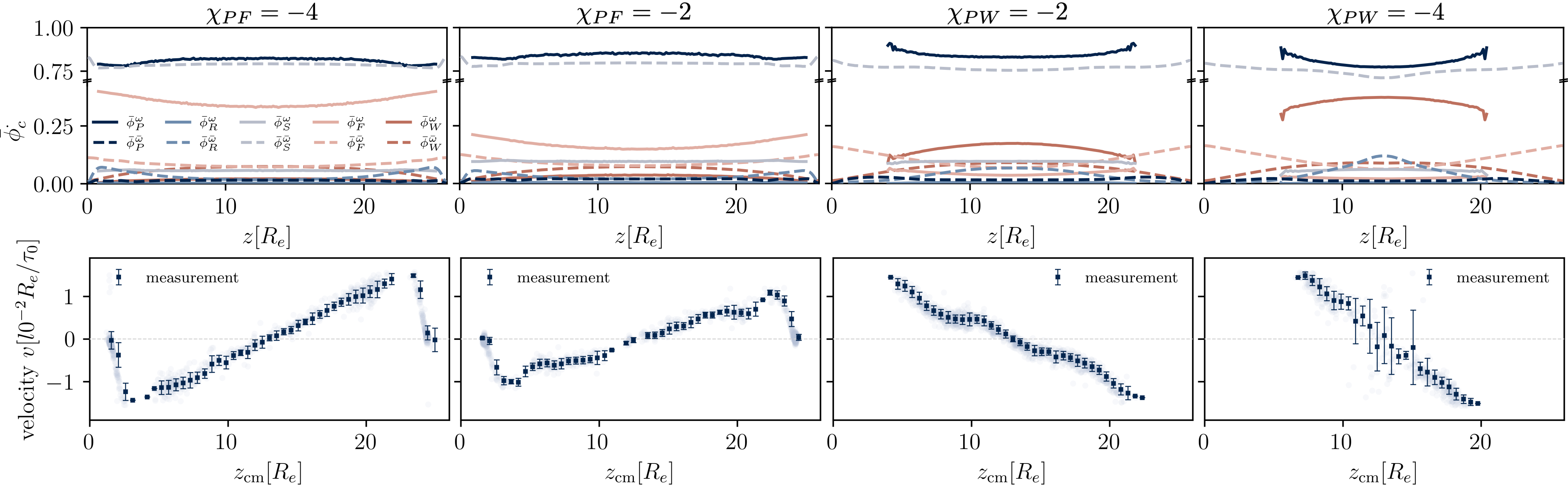}
    \caption{Results of the particle-based simulations for reaction rate constant $r_b=0.08\tau_0^{-1}$ and variation of interaction parameters. (a) Domain-averaged densities for selected components. (b) Measured individual (light blue circles) and averaged (blue squares) velocities.}
    \label{fig:soma_results}
\end{figure}

\section{Simulations of Multiple Droplet Types}
\label{sec:parameters-multiple-droplet-types}
The system presented in \autoref{sec:multiple-droplet-types} features two droplet molecule types, $P$ and $Q$, which phase separate into two distinct droplet phases. Here, we present the parameters that are used and how the simulations are set up.  We take a simulation from \autoref{sec:explicit-fuel-waste} where the product droplets are attractive towards fuel, $\chi_{PR}N_P=20,\chi_{PS}N_P=50,\chi_{PF}N_P=-20,\chi_{PW}N_P=0$. Using the final state of that simulation as input, we replace part of the solvent in the system by another homopolymer, $Q$, such that it makes up $5\%$ of the system's volume. The new component phase separates but favorably interacts with the waste, $\chi_{QR}N_P=20,\chi_{QS}N_P=50,\chi_{QF}N_P=0,\chi_{QW}N_P=-20$. In order to observe two types of droplets that do not wet each other, we take a strong repulsion of $P$ and $Q$ polymers, $\chi_{PQ}N_P=200$. We show that upon demixing from solution, the $Q$-rich droplets move towards the center, whereas the $P$-rich droplets continue to move towards the fuel source. This is demonstrated in \autoref{fig:multidroplets} of the main text, where example snapshots and velocity measurements demonstrate that even the $Q$-rich droplets whose components are not involved in the reaction cycle, exhibit a directed movement.

\section{Simulations of Amphiphilic Product Molecules}
\label{sec:parameters_amphiphiles}

When introducing one component into the system that is amphiphilic, we need to extend our continuum model to include this. Modeling the amphiphiles with a theory for diblock copolymers, we employ the free-energy functional proposed by Uneyama and Doi \cite{uneyama_density_2005,uneyama_calculation_2005,uneyama_density_2007}. The amphiphilic diblock copolymer consists of two linear blocks with distinct monomer types, called $A$ and $B$. For each of these blocks, we introduce a density field, as we do for a other single-block components, \ie, $R,S,F,W$. Introducing for molecular type index $p$ and for the blocks of each molecule index $i$, the free-energy functional reads
\begin{eqnarray}
  \frac{\mathcal{F}\left[\left\{\phi_{p i}(\mathbf{r})\right\}\right]}{\sqrt{\Nbar}\kT}
  &=& \frac{1}{\R^3}\int \mathrm{d}\mathbf{r} \left[\pi(\mathbf{r})N_P\left(\sum_{pi} \phi_{pi} - 1 \right)\right. \nonumber\\
  &&+\sum_{p, i j} \int  \, \mathrm{d} \mathbf{r}^{\prime}\, 2 \sqrt{f_{p i} f_{p j}} A_{p, i j} \mathcal{G}(\boldsymbol{r}-\mathbf{r}^{\prime}) \sqrt{\phi_{p i}(\mathbf{r}) \phi_{p j}\left(\mathbf{r}^{\prime}\right)} \nonumber  \\
  &&+\sum_{p i} f_{p i} C_{p, i i} \phi_{p i}(\mathbf{r}) \ln \phi_{p i}(\mathbf{r})\nonumber \\
  &&+\sum_{p, i \neq j}  2 \sqrt{f_{p i} f_{p j}}  C_{p, i j} \sqrt{\phi_{p i}(\mathbf{r}) \phi_{p j}(\mathbf{r})} \nonumber \\ 
  &&+\sum_{p i} \frac{\R^2}{12\phi_{p i}(\mathbf{r})}  \left|\nabla \phi_{p i}(\mathbf{r})\right|^{2} \nonumber \\
  &&+\left.\frac{1}{2}\sum_{p i, q j} \chi_{p i, q j}N_P \phi_{p i}(\mathbf{r}) \phi_{q j}(\mathbf{r}) \right]
  \label{eq:UDM_general} 
\end{eqnarray}
where the coefficients $A{p,ij}$ and $C_{pij}$ and the parameters $f_{pi}$ are dependent on the molecular architecture. For simple homopolymers and solvents, the theory reduces to the Flory-Huggins-de Gennes model \cite{de_gennes_dynamics_1980} as introduced in the main text, $A_p=0$, $C_p=N_P/N_p$ and $f_p=1$. For the diblock copolymer, the free-energy expansion to second order in the concentrations matches Ohta-Kawasaki free-energy \cite{ohta_equilibrium_1986}. For these components, the coefficients become \cite{hafner_reaction-driven_2023,dreyer_simulation_2022}
\begin{eqnarray}
  A_{p} &=& \frac{9N_P^2}{N_p^2\R^2f_{\rm{A}}^2f_{\rm{B}}^2} \left(\begin{matrix}
  	f_{\rm{B}}^2 & -f_{\rm{A}}f_{\rm{B}} \\ -f_{\rm{A}} f_{\rm{B}} & f_{\rm{A}}^2
  \end{matrix} \right) \\
 C_{p} &=& \frac{N_P}{N_p f_{\rm{A}} f_{\rm{B}}} \left( \begin{matrix}
  	\tilde s(f_{\rm{A}}) & -\frac{1}{4} \\ -\frac{1}{4} & \tilde s(f_{\rm{B}}) 
  \end{matrix}\right)  
\end{eqnarray}
where $f_{\rm B}=1-f_{\rm A}=0.7$ is the block ratio and the definition 
\begin{align}
     \tilde s(f) = \frac{1}{4} \left( \frac{s(f)}{f(1-f)} - \frac{1}{1-f} \right).
  \label{eq:stilde}
\end{align}
The reactions are implemented as described in Ref.~\cite{hafner_reaction-driven_2023} which assumes a single reaction to convert the hydrophilic precursor into an amphiphilic molecule. This would correspond to short molecules with presumably weak amphiphilicity, where a single reaction can create such a drastic change in the molecule's interactions. Overall, we still describe the reaction cycle by a forward and backward rate constant, $r_f$ and $r_b$, respectively.
The system consists finally of a hydrophilic precursor, $R$, in solution, $S$, that reacts with a fuel molecule, $F$, to become amphiphilic, \ie, the tail, $B$, becomes hydrophobic, while the head group, $A$, stays hydrophilic. Waste, $W$, is produced within the forward reaction and the backward reaction from amphiphile to precursor happens spontaneously, modeled as a first order reaction. The chain discretizations are taken as in the main text, the amphiphilic molecules have a block ratio of $f_A=1-f_B=0.3$, and we take the interaction matrix

\begin{align}
&\quad\,\,{\color{gray} A\,\,\,\, \,B\,\,\,\, \,R\,\,\,\, \,S\,\,\, \,\,F\,\,\,\,\, W} \nonumber\\  
\chi_{cc'} N_P = &\left(\begin{matrix}
    0  & 20 & 0 & -5 & -5 & 5 \\
    20 & 0  & 20& 50 & 50 & 50 \\
    0  & 20 & 0 & 0  & 0  & 0  \\
    -5 & 50 & 0 & 0  & 0  & 0  \\
    -5 & 50 & 0 & 0  & 0  & 0  \\
     5 & 50 & 0 & 0  & 0  & 0  
\end{matrix}\right){\color{gray}\begin{array}{c}
    A  \\
    B \\
    R\\
    S\\
    F\\
    W
\end{array}}
\end{align}
We take larger simulation-cell sizes of $V=51.2\R\times 9.6\R\times 9.6\R$ such that micelles do not grow too quickly after nucleation, but slowly grow upon moving from the bulk towards the source, since the head-groups interact favorably with the fuel. We take a backward reaction rate of $r_b=0.04\lambda$ and refilling rates of $r_{\mathrm{source}}=0.02\lambda\R$, $r_{\mathrm{sink}}=10\lambda\R$. The result is a movement of aggregates as depicted in \autoref{sec:complex-assemblies}, where the slow growth allows the assembly of vesicles, that would not form if it the amphiphiles would just experience a quench into the phase separating parameter regime. 

Indeed, performing simulations without fuel and waste and with either no reactions or first-order reactions between precursor and product, but taking all interactions to be the same, we observe no coarsening after the quench and initial nucleation of small micelles, see \autoref{fig:amphiphile-stationary-test}. This also becomes visible when plotting micelle radii against time, where the sizes just stall at the initially unstable length scale and no fusion happens. 

\begin{figure}
    \centering
    \includegraphics[width=0.5\textwidth]{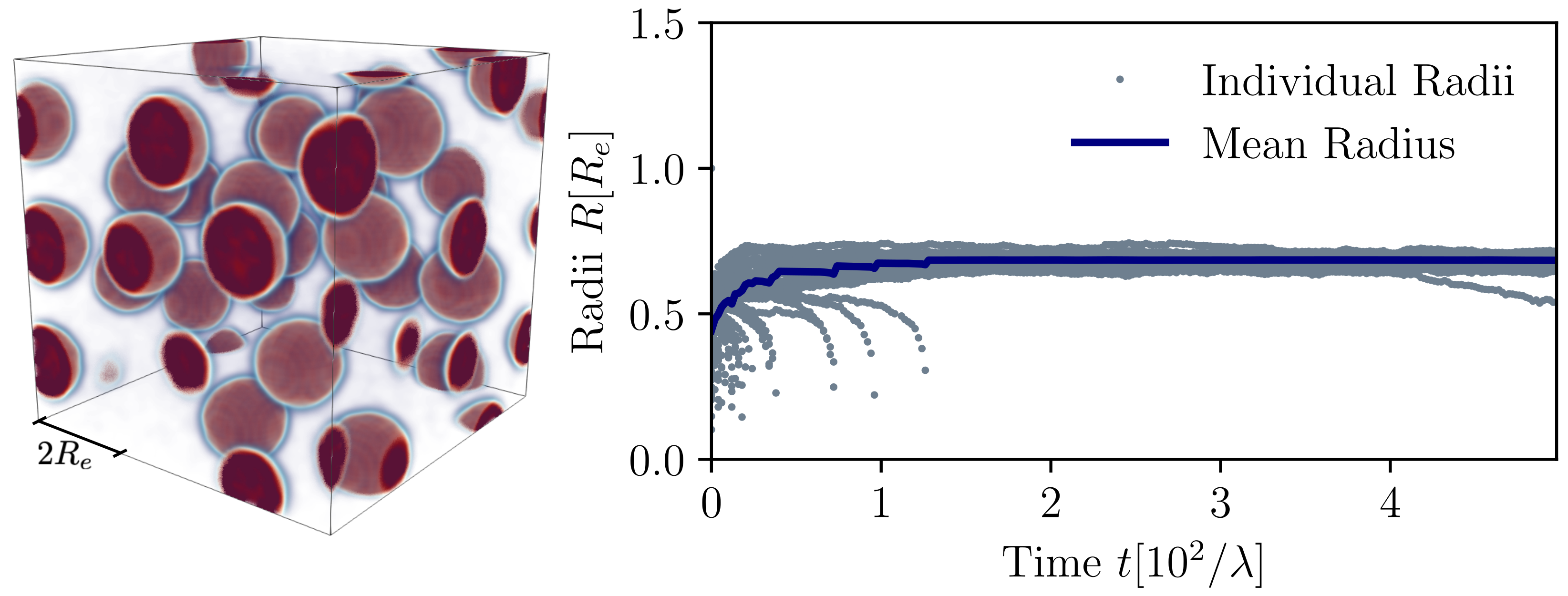}
    \caption{Simple quench test of amphiphiles in equilibrium with the interactions kept the same. Left: Morphology of tail-groups, $\phi_B(\mathbf{r})$ after time $t\lambda=500$. Right: Time evolution of micelle radii.}
    \label{fig:amphiphile-stationary-test}
\end{figure}

\section{Analysis Procedure}
\label{sec:analysis_procedure}

The evaluation of the theoretical prediction of \autoref{sec:velocities_analytical} requires proper analysis of the simulation data because we use the concentration and flux measurements to compare these to the velocity measurements. Here, we describe step-by-step what is necessary for the simulations and to analyze the (noisy) simulation results.

\paragraph{Simulation}
As of our implementation of the Uneyama-Doi model (UDM) (available open source \textit{via} \url{www.gitlab.com/g.ibbeken/udm}), the simulation is a two-step procedure:
\begin{enumerate}
    \item An \ac{XML} input file is used specify the system architecture (including simulation cell size, grid spacing, external field) the different molecular components (including molecular volume and their initial mean concentrations), as well as the reaction dynamics. The software provides a script to create a \ac{HDF5} file which is the input to the simulation.
    \item The \ac{HDF5}-file is given to the simulations which requires the presence of a \ac{CUDA}-\ac{GPU}. The concentration fields are regularly saved according to an interval, $\Delta \tau$, as specified in the \ac{XML} input file.
\end{enumerate}

The \ac{HDF5} file now contains the simulation results in the form of the time-dependent ($n_t$ saved times) density fields for all $n_c$ components. In the file a $n_t\times n_c\times n_z\times n_y\times n_x$ sized array is saved.

\paragraph{Analysis}
The analysis takes the time-dependent data and performs the following steps
\begin{enumerate}
    \item Perform a \ac{HKCA} on all time-steps, defining a cluster where $\phi_P>0.5$. The result, a cluster map, is saved in the \ac{HDF5}-file as a $n_t\times n_z\times n_y\times n_x$-sized array which is zero if no cluster is detected and non-zero, indicating the cluster index, otherwise.
    \item Using the \ac{HKCA}, calculate the cluster radii, $R$, as described in Ref. \cite{hafner_reaction-driven_2023} and the center of mass, $\mathbf{r}_{\mathrm{cm}}$. 
    \item Using the center of mass, compute the Voronoi tessellation by determining for each grid cell the cluster with the closest center of mass. The result is stored to the \ac{HDF5}-file.
    \item If it is of interest, the Voronoi-domains can be analysed in detail. For this the domain is separated into $51$ bins (bin width of $\sim 0.5\R$). Within each bin, we can averaged each Voronoi domain within the center-of-mass frame. From this average, the averaged volume in front of ($z<z_{\mathrm{cm}}$) or behind ($z>z_{\mathrm{cm}}$) the droplet can be computed.
    \item Using the \ac{HKCA}, the concentration fields can be averaged in- and outside of the droplet phases. The cluster analysis separates the droplet from its outside exactly at the interface. Hence, simply taking averaged concentration where the cluster map is non-zero results in serious miss-evaluation of the inside concentrations of droplets, if these are small (high surface-to-volume ratio). Instead we refine the cluster, making them smaller by excluding grid cells that are (next-nearest-)neighbors of non-cluster grid cells. Averaging over the refined clusters now excludes the interface, leading to clean concentration profiles. For the analysis of the outside concentration profiles the opposite refinement is performed. 
    \item The velocities are measured using the center-of-mass measurement. For this, the same droplet within two time-frames are detected by finding for each cluster (droplet) a cluster with the closest center of mass (with distance smaller than $|\mathbf{r}_{\mathrm{cm},i,t}-\mathbf{r}_{\mathrm{cm},i',t+\Delta\tau}|\stackrel{!}{<}\R$ for cluster $i$ at time $t$ and cluster $i'$ at time $t+\Delta \tau$). The difference in center of mass then determines the velocity.
    \item Using \autoref{eq:velocity_symmetric_final}, the predictions are calculated using the above analysis. To obtain the flux inside the droplets, the noisy data has to be filtered (to obtain clean gradients). To this end, we run the continuum model simulation without thermal noise for a short time, $t\lambda=0.1$ which is long enough to properly filter the fluctuations and short enough to not alter the concentration fields significantly. The fluxes inside the droplets are measured using the cluster analysis. The reaction contribution uses the (smoothed) local concentration fields in combination with the Voronoi tessellation.

\end{enumerate}
\end{widetext}
\end{document}